 \documentclass{emulateapj}

\newcommand{\sersic}{S\'ersic}
\newcommand{\HST}{{\em HST}}
\newcommand{\ACS}{{\em ACS}}
\newcommand{\combo}{{COMBO-17}}
\newcommand{\gems}{{GEMS}}
\newcommand{\sdss}{SDSS}
\newcommand{\goods}{GOODS}
\newcommand{\galfit}{{\sc galfit}}
\newcommand{\gimtwod}{{\sc gim2d}}
\newcommand{\sex}{{\sc SExtractor}}
\newcommand{\reapp}{R_e^{\textrm{\tiny{app}}}}
\newcommand{\rtot}{m_{\textrm{\tiny{R,tot}}}}
\newcommand{\raper}{m_{\textrm{\tiny{R,aper}}}}
\newcommand{\logm}{\log\left(\mathcal{M}/\mathcal{M}_\odot\right)}
\newcommand{\muav}{\left\langle\mu_V\left(z\right)\right\rangle}
\newcommand{\rhoav}{\left\langle\log\Sigma_\mathcal{M}\left(z\right)\right\rangle}

\slugcomment{{\sc Draft}{\it: February 21, 2005}}

\shorttitle{\gems: The Size Evolution of Disk Galaxies}
\shortauthors{Barden et al.}

\begin{document}

\title{\gems: The Size Evolution of Disk Galaxies}

\author{
Marco Barden\altaffilmark{1}, 
Hans-Walter Rix\altaffilmark{1}, 
Rachel S.~Somerville\altaffilmark{2}, 
Eric F.~Bell\altaffilmark{1}, 
Boris H\"au\ss ler\altaffilmark{1}, 
Chien Y.~Peng\altaffilmark{3,2}, 
Andrea Borch\altaffilmark{1}, 
Steven V.~W.~Beckwith\altaffilmark{2,4}, 
John A.~R.~Caldwell\altaffilmark{2,9}, 
Catherine Heymans\altaffilmark{1}, 
Knud Jahnke\altaffilmark{5}, 
Shardha Jogee\altaffilmark{2,10}, 
Daniel H.~McIntosh\altaffilmark{6}, 
Klaus Meisenheimer\altaffilmark{1}, 
Sebastian F.~S\'anchez\altaffilmark{5,11}, 
Lutz Wisotzki\altaffilmark{5,7}, 
Chrisitan Wolf\altaffilmark{8}
}
\email{barden@mpia.de}

\altaffiltext{1}{Max-Planck-Institut f\"ur Astronomie, K\"onigstuhl 17, 69117, Heidelberg, Germany}
\altaffiltext{2}{Space Telescope Science Institute, 3700 San Martin Dr., Baltimore, MD 21218, USA}
\altaffiltext{3}{Steward Observatory, University of Arizona, 933 N.\ Cherry Ave., Tucson AZ, 85721, USA}
\altaffiltext{4}{Johns Hopkins University, 3400 North Charles Street, Baltimore, MD 21218, USA}
\altaffiltext{5}{Astrophysikalisches Institut Potsdam, An der Sternwarte 16, 14482, Potsdam, Germany}
\altaffiltext{6}{Department of Astronomy, University of Massachusetts, 710 North Pleasant Street, Amherst, MA 01003, USA}
\altaffiltext{7}{Universit\"at Potsdam, Am Neuen Palais 10, 14469, Potsdam, Germany}
\altaffiltext{8}{Department of Physics, Denys Wilkinson Bldg., University of Oxford, Keble Road, Oxford, OX1 3RH, UK}
\altaffiltext{9}{Present address: University of Texas, McDonald Observatory, Fort Davis, TX 79734, USA}
\altaffiltext{10}{Present address: University of Texas at Austin, 1, University Station C1400, Austin, TX 78712-0259, USA}
\altaffiltext{11}{Present address: Centro Astronomico Hispano Aleman de Calar Alto, C/Jesus Durban Remon 2-2, Almeria, E-04004, Spain}

\begin{abstract}
We combine HST imaging from the \gems\footnote{Galaxy Evolution from
  Morphologies and SEDs} survey with photometric redshifts from
\combo\ to explore the evolution of disk-dominated galaxies since $z
\la 1.1$.  The sample is comprised of all \gems\ galaxies with
\sersic\ indices $n<2.5$, derived from fits to the galaxy images. We
account fully for selection effects through careful analysis of image
simulations; we are limited by the depth of the redshift and HST data
to the study of galaxies with $M_V\lesssim-20$, or equivalently
$\logm\gtrsim 10$.  We find strong evolution in the magnitude--size
scaling relation for galaxies with $M_V\lesssim-20$, corresponding to
a brightening of $\sim$1 mag\,arcsec$^{-2}$ in rest-frame $V$-band by
$z\sim 1$.  Yet, disks at a given absolute magnitude are bluer and
have lower stellar mass-to-light ratios at $z\sim 1$ than at the
present day.  As a result, our findings indicate weak or no evolution
in the relation between stellar mass and effective disk size for
galaxies with $\logm\gtrsim 10$ over the same time interval.  This is
strongly inconsistent with the most naive theoretical expectation, in
which disk size scales in proportion to the halo virial radius, which
would predict that disks are a factor of two denser at fixed mass at
$z\sim1$. The lack of evolution in the stellar mass--size relation is
consistent with an ``inside-out'' growth of galaxy disks on average
(galaxies increasing in size as they grow more massive), although we
cannot rule out more complex evolutionary scenarios.

\end{abstract}


\section{Introduction}

The last eight billion years have witnessed strong evolution of
the disk galaxy population.  Both `archaeological' studies
of local disk-dominated galaxies 
and `look-back' studies of the evolution of disk galaxies
suggest a steady build-up in their stellar masses
since $z \sim 1$ \protect\citep{2000A&A...358..869R,1999ApJ...517..148F,
astro-ph/0502246,astro-ph/0410518}.  Insights into how this growth occurs
are accessible through the study of disk galaxy scaling
relations, such as the luminosity--rotation 
velocity (Tully-Fisher) relation or the luminosity--size
relation \protect\citep[e.g.][]{1996ApJ...465L..15V,1998ApJ...500...75L,
1999ApJ...519..563S,2004A&A...420...97B}.  Yet, 
owing to sample size limitations, selection effects, 
and differences in analysis techniques, these 
studies have come to widely divergent conclusions.
In this paper, we explore the evolution 
of the luminosity--size and stellar mass--size relations
over the last 8 Gyr (since $z \sim 1$) using a sample of almost
5700 disk-dominated galaxies from the HST GEMS survey \protect\citep[Galaxy 
Evolution from Morphology and SEDs][]{2004ApJS..152..163R}.

In the Cold Dark Matter (CDM) picture of structure formation, collapsing
dark matter perturbations acquire angular momentum through tidal torques
and mergers \protect\citep{peebles:69,mds,vitvitska}.  Some fraction 
of this angular momentum is conserved, leading to the formation of
cold, rotationally-supported disks.  The typical
magnitude of the specific angular momentum predicted in this framework
leads to the formation of present day disks with approximately the correct
distribution of radial sizes, if the specific angular momentum of
the gas is similar to that of the dark matter and is mostly 
conserved during the formation process \protect\citep{fall_efstathiou}.

A difficulty is that this idealized picture does not
correspond to the outcome when the process of galaxy formation is
simulated in detail within the cosmological context of CDM. In
hydrodynamical simulations, the gas tends to lose a large fraction of its
initial angular momentum, resulting in disks that are too small 
compared to observed nearby galaxies 
\protect\citep{nw:94,sgv:99,ns:00,2004ApJ...612L..13D}.
Furthermore, very few `late-type' disks
are formed in such simulations: galaxies tend to suffer mergers that
thicken and destroy their disks \protect\citep{sn:02}. It is not yet established
whether this problem represents a fundamental difficulty with the
`standard' CDM paradigm (i.e., a result of excess small scale power), a
reflection of our incomplete ability to understand and simulate the
complexities of star formation and supernova feedback, or 
inadequacies in numerical resolution. 

Many proposed solutions to this problem involve \emph{delaying} gas
collapse and disk formation to later times, either by adopting an
alternate power spectrum with reduced small scale power (such as Warm Dark
Matter), in which structure formation occurs later \protect\citep[e.g.][]{sld:01},
or by invoking some form of feedback that prevents the gas from
cooling until relatively late times $z\sim1$ \protect\citep{weil:98,tc:01}.
While these solutions would be consistent with an important 
build-up in the disk galaxy population at late times, the late formation times
implied by these models may be in conflict
with the old ages of disk stars in the Milky Way and M31 
\protect\citep{2000A&A...358..869R,fj:01}.  Additional
constraints can be gleaned from so-called `backwards evolution' models, 
in which the ages and metallicities of the stars in present-day disk 
galaxies are used to constrain the formation history of different
components within our and other galaxies
\protect\citep{1997ApJ...477..765C,1999MNRAS.307..857B}.  Direct measurements of
the mass--size scaling relations and radial size distributions of disk
galaxies at earlier epochs will provide an important counterpoint to these
arguments by providing direct constraints on the angular momentum content
of stars at these earlier times.

A number of previous studies have used the 
{\em Hubble Space Telescope} (\HST) to quantify the
evolution of disk galaxies by measuring their absolute sizes and
magnitudes as a function of redshift.  Magnitude and size are
strongly correlated; a line of constant surface brightness 
falls {\it almost} parallel to the
distribution of observed galaxies, making the evolution 
of galaxy surface brightness a natural choice for
parameterizing the evolution of galaxy sizes. 
However, the results of studies
measuring average rest-frame surface brightnesses as a function of
redshift have proven controversial, ranging from detecting no evolution to
rather strong evolution in the range of 1-2 mag arcsec$^{-2}$ brightening
by redshift $z\sim 1$. For example, \protect\citet{1998ApJ...500...75L} found an
average increase of the surface brightness of $\sim 1$ mag by redshift
$z\sim 1$. This result is supported by observations of galaxies at high
redshifts ($z\sim2-3$), detected in very deep ground-based near-infrared
images \protect\citep{2003ApJ...591L..95L}. \protect\cite{2004ApJ...604..521T} estimate
that the average rest-frame surface brightness of these objects is more
than 2-3 mag arcsec$^{-2}$ brighter than in the local universe.

\protect\citet{1999ApJ...519..563S} pointed out that selection
effects play a crucial role in such analyses.  After accounting for 
the different sources of incompleteness, \protect\citet{1999ApJ...519..563S} 
and \protect\citet{2004ApJ...604L...9R} argue that the luminosity--size
relation of disk galaxies evolves by less than 0.4 mag arcsec$^{-2}$ 
over the interval $0.25<z<1.25$.  Yet, in order to reproduce the observations,
both groups found it necessary to introduce a 
new population of high surface brightness
galaxies in the highest redshift bin ($z\sim 1$). 
A different interpretation was suggested by \protect\citet{2004MNRAS.355...82T}, who
find strong evolution of the average rest-frame $V$-band surface 
brightness of $\sim0.8$ mag arcsec$^{-2}$ at a redshift $z\sim0.7$, 
also including a full treatment of completeness.

In this work, we present the results from a new sample of disk-dominated
galaxies from the \gems\ survey. Each of our galaxies has a
spectrophotometrically-measured redshift, a spectral energy distribution 
\protect\citep[SED]{2004A&A...421..913W}, and a stellar mass estimate 
\protect\citep{2004PhDT.........3B} from \combo\ .  We combine these SED constraints
with light-profile shapes and sizes determined from deep high-resolution
\HST\ {\em Advanced Camera for Surveys} (\ACS) images. We reassess the
evolution of the magnitude--size and stellar mass--size relation as a
function of redshift over the range $0.1\lesssim z\lesssim 1.1$, 
taking particular care
to model the impact of the selection function. We suggest a resolution to
the conflicting previous results by presenting a coherent picture of
strong surface brightness evolution with redshift without the need for a
new population of high surface brightness galaxies. 

The layout of this paper is as follows. In $\S$~\ref{sec_the_sample}
we present the \gems\ data set and describe the sample selection, the
galaxy fitting techniques and the corrections we applied to the
data. We explain in some more detail our modeling of the sample
completeness in $\S$~\ref{sec_analysis}. In
$\S$~\ref{sec_mag_mass_size}, we explore the evolution of the
magnitude--size and stellar mass--size relations for disk-dominated
galaxies. We show that there is a trend of increasing average surface
brightness with redshift and that there is little evolution of the
surface mass density.  In $\S$~\ref{discussion} we discuss our results
in comparison with previous studies in the literature, and compare
them with theoretical expectations.  We summarize our results in
$\S$~\ref{sec_summary}.  Throughout this paper we use the concordance
cosmology with $H_0=70$ km s$^{-1}$ Mpc$^{-1}$, $\Omega_M=0.3$ and
$\Omega_\Lambda =0.7$ \protect\citep{2003ApJS..148..175S}. Unless indicated
otherwise we use Vega-normalized magnitudes.

\section{Sample Definition}\label{sec_the_sample}

\subsection{Imaging Data}

\gems, Galaxy Evolution from Morphologies and SEDs
\protect\citep{2004ApJS..152..163R}, has imaged an area of $\sim 800$ arcmin$^2$
centred on the Chandra Deep Field South (CDFS), using the \ACS\ on-board
\HST. Of these 78 \ACS\ tiles the central 15 were incorporated from
the \goods\ project \protect\citep{2004ApJ...600L..93G}. With integration times of
$\sim 35$ min in each of two filters (F606W and F850LP) the point source
detection limits reached
$m_\textrm{\tiny{AB}}\left(\textrm{F606W}\right)=28.3$ ($5\sigma$) and
$m_\textrm{\tiny{AB}}\left(\textrm{F850LP}\right)=27.1$ ($5\sigma$),
respectively. Details about the image mosaic and data reduction will be
explained in a subsequent paper (Caldwell et al.~2005, in prep.).

\subsection{COMBO-17 Data}

The \HST\ imaging data is complemented by low resolution
spectrophotometric data from
\combo\ \protect\citep{2004A&A...421..913W}. \combo\ has provided precise
redshift estimates ($\sigma_z/(1+z)\sim 0.02$) for approximately 9000
galaxies down to $m_R<24$.  Rest-frame absolute magnitudes and colors,
accurate to $\sim$0.1\,mag, are also available for these galaxies.
Furthermore, using a simple parameterized star formation history and
the photometry in the 17 \combo\ bands, \protect\citet{2004PhDT.........3B}
computed stellar mass estimates for each galaxy in our sample,
assuming a \protect\citet{1993MNRAS.262..545K} stellar initial mass function
(IMF).  These mass estimates are consistent with those derived
using a one-color-based transformation from light to mass as described
in \protect\citet{2001ApJ...550..212B} and \protect\citet{2003ApJS..149..289B}.  While
such estimates suffer from uncertainties in the IMF, ages, dust,
and metallicity, it is encouraging to note that several studies
\protect\citep{2003ApJS..149..289B,2004ApJ...616L.103D} find good agreement
between masses based on broad-band colors and those from spectroscopic
\protect\citep[e.g.][]{2003MNRAS.341...33K,2003MNRAS.341...54K} and dynamical
\protect\citep{2004ApJ...616L.103D} techniques.

\subsection{Source Detection}

For source detection we use the \sex\ software \protect\citep{1996A&AS..117..393B}
on the F850LP image. In contrast to the standard single-pass
approach, we apply a two-step process, running \sex\ twice on each image
to find an acceptable compromise between deblending and detection
threshold \protect\citep[see][]{2004ApJS..152..163R}. Combining the source lists
from each tile, taking care to remove duplicate objects that were detected
in two neighbouring tiles, we end up with over 40,000 galaxies.

\subsection{Galaxy Fitting and Disk Selection}\label{selection}

For the purpose of this paper we wish to isolate the subset of galaxies
whose light is dominated by a disk component. We start by identifying all
galaxies that can be reasonably well-fit by any single \sersic\ profile
\protect\citep{sersicprofile} using the two-dimensional fitting code \galfit\
\protect\citep{2002AJ....124..266P}. The \sersic\ profile is a generalisation of a
de Vaucouleurs profile with variable \sersic\ index $n$: 
\begin{equation}
\Sigma\left(R\right)=\Sigma_e\times\exp\left(-\kappa\left[
\left(R/R_e\right)^{1/n}-1\right]\right), 
\end{equation} 
where $R_e$ is
the effective or half-light radius, $\Sigma_e$ is the effective surface
density, $\Sigma\left(R\right)$ is the surface density as a function of
radius and $\kappa=\kappa\left(n\right)$ is a normalization constant. 
An exponential profile has $n=1$ while a de Vaucouleurs profile has
$n=4$. \galfit\ convolves \sersic\ profile galaxy models with the point
spread function of the \ACS\ 
\protect\citep[Jahnke et al., in preparation]{2004ApJ...614..568J} and
then determines the best fit by comparing the convolved models with the
science data using a Levenberg-Marquardt algorithm to minimize the
$\chi^2$ of the fit. The best-fit model is given by 7 parameter values and
their associated uncertainties, including the half-light radius, the
\sersic\ index and the total magnitude. Initial \galfit\ starting guesses
for the model parameters were obtained from the \sex\ source catalogues.
Typically, neighbouring galaxies were excluded from each model fit using
a mask, but in the case of closely neighbouring galaxies with
overlapping isophotes the galaxies were fitted simultaneously. The sky
level for each galaxy was carefully measured using flux growth curves,
masking out detected neighbouring sources. Lacking an estimate for the
\sersic\ index from \sex, we started all fits with $n=1.5$. In addition, 
all galaxies with $0.65<z<0.75$ were fitted with \gimtwod\
\protect\citep{2002ApJS..142....1S}. Estimates for magnitudes, sizes and \sersic\ 
indices from the two codes agree very well (see Bell et al.\ 2004; 
H\"au\ss ler et al.~2005, in prep.). 
Morphological quantities quoted in the present paper were
derived using \galfit.

For this study, we estimate structural and morphological parameters from
the $z$-band images (F850LP). In
the optical (and the near-infrared), young
stars make a progressively smaller contribution with 
increasing wavelength.  Therefore, galaxy morphologies in
F850LP are smoother than those in F606W, leading to a more 
robust detection and deblending of extended sources.  
The F850LP band corresponds to rest-frame $R$, $V$, and 
$B$-bands at $z \sim 0.4$, 0.7, and 1 respectively.

\begin{figure}
\figurenum{1}
\begin{center}
\includegraphics[width=3.25in,keepaspectratio,clip]{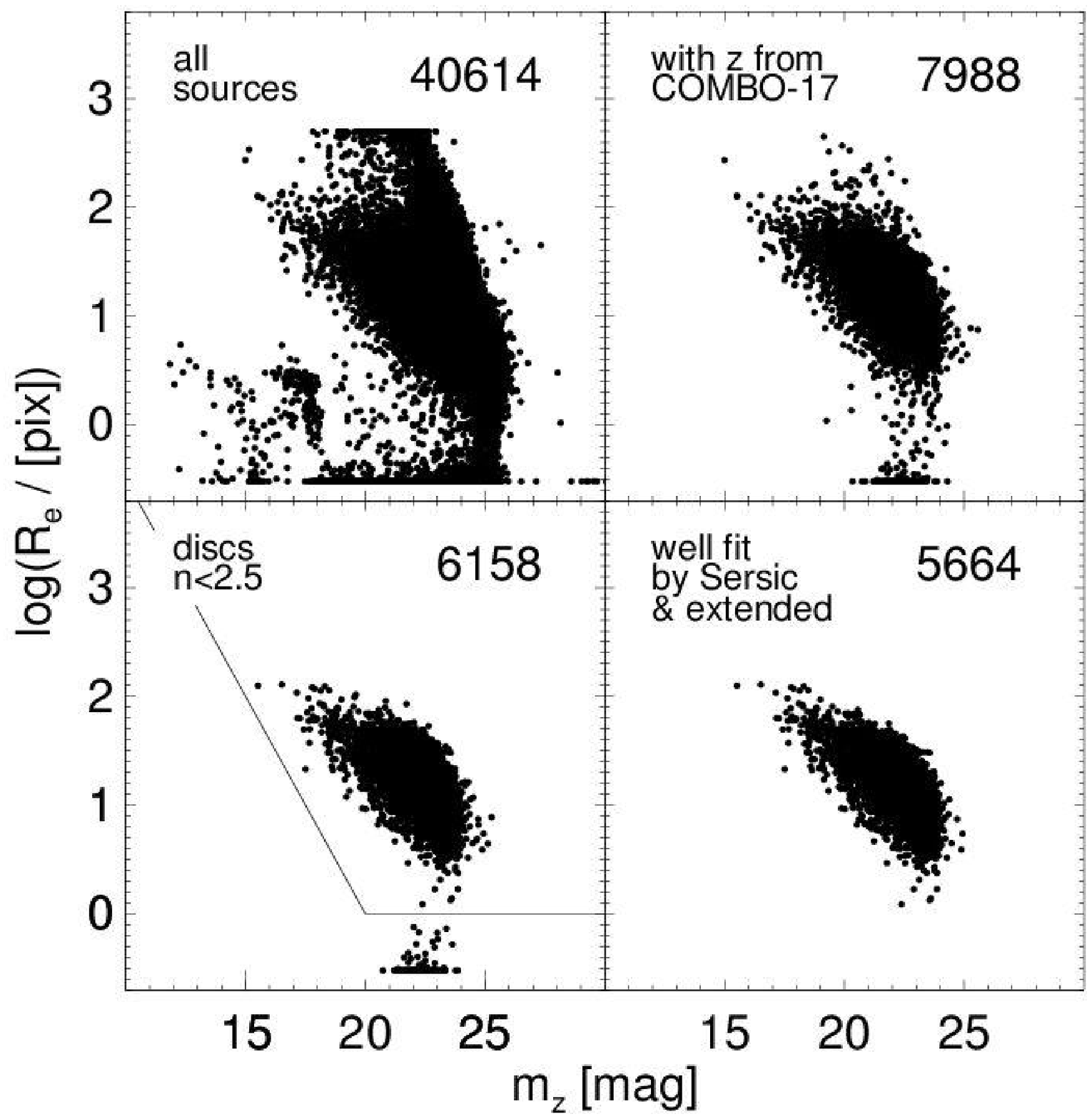}
\end{center}
\caption{The disk sample definition in the apparent magnitude--size plane
(apparent effective radius: $\reapp$, apparent $z$-band
magnitude: $m_z$). Top left: all galaxies detected in the \gems\ tiles.
Top right: galaxies with \combo\ redshifts. Bottom left: disk galaxies 
with \sersic\ index $n<2.5$. Bottom right: disk galaxies with reliable 
\galfit\ fits (see text for details on selection criteria). In each 
panel, we give the total number of galaxies in the upper right.
\label{fig_sample}}
\end{figure}

Selection of a galaxy sample for this kind of study is a multi-step
process. First we merge the \gems\ catalogue with the \combo\ redshift
catalogue, then we select disk-dominated objects, and finally we remove
sources with poor fits (see Fig.~\ref{fig_sample}). We start by
matching the \gems\ sources to the \combo\ catalogue. To account for
the relatively high source density in the \gems\ images we pick the
closest neighbour in the \combo\ catalogue within 0.5 arcsec as the
corresponding match for a \gems\ galaxy. Only at matching distances
exceeding 1 arcsec does one start to include uncorrelated pairs. We
are left with about 8000 matched sources in our sample.

We isolate disk-dominated galaxies for further study by cutting the sample
based on the \sersic\ profile fits.  We adopt
$n=2.5$ as the dividing line between disk- ($n<2.5$) and
spheroid-dominated ($n>2.5$) galaxies.
This cut discriminated between visually-classified 
early- and late-type galaxies from GEMS with 
$0.65 < z < 0.75$ with 80\% reliability and less than 25\% contamination
\protect\citep{2004ApJ...600L..11B}.   This cut is also 
consistent with the analysis
conducted by the Sloan Digital Sky Survey \protect\citep[\sdss;
see][]{2003MNRAS.343..978S}. 
Furthermore, \protect\citet{2004ApJ...604L...9R} have redshifted a sample of local
galaxies to show that the \sersic\ index is still a useful indicator at
redshifts $z\sim 0.5-1$. Selecting galaxies with $n<2.5$ left us with
$\sim6200$ disk-dominated objects.

To ensure that the extracted galaxy profile parameters are reliable we
remove objects from our source list that have relative formal errors in \sersic\ 
index $n$ and effective radius $R_e$ of more than 25\% ($\delta
n/n>0.25$, $\delta R_e/R_e>0.25$)\footnote{\galfit\ formal 
errors underestimate the true uncertainties, as assessed using 
simulated galaxy images.  The true uncertainties for the bulk
of the sample are $\sim 35$\% in $R_e$ and $\sim 0.2$\,mag in $m_z$.}. 
We also exclude objects that reach the boundary conditions for
$n$ ($0.2<n<8$) or $R_e$ ($0.3<R_e$ [pixel] $<500$). Furthermore, we require
that the \galfit\ magnitudes coincide with the \sex\ magnitudes to within
0.6 mag
($|m_{\textrm{\tiny{GALFIT}}}-m_{\textrm{\tiny{SEx}}}+0.166|<0.6$).
Finally, we remove compact sources with 
$\log\reapp<\textrm{max}\left[8-0.4\times m_z,0\right]$ (indicated by the 
solid line in the bottom left panel in Fig.~\ref{fig_sample}).
While slightly more galaxies with low surface brightness were removed
by these additional cuts than high surface brightness galaxies, no 
pronounced bias was introduced.  It is important to note that 
the simulated galaxy samples were also subjected to these 
same cuts for the construction of the completeness maps; thus, the
completeness maps account fully for any biases introduced by these
(necessary) extra sample cuts.

This sample selection should provide a fair representation of the
disk-dominated galaxy population at all redshifts. The final
catalogue contains 5664 disk galaxies with absolute rest-frame $B$- and
$V$-band magnitudes, redshifts and stellar masses obtained from \combo\
and apparent half-light radii and \sersic\ indices from \galfit.

\subsection{The Local Comparison Sample}\label{sdss}

In order to compare our measurements to a local reference point we have 
opted to use the NYU Value-Added Galaxy Catalog 
\protect\citep[VAGC;][]{astro-ph/0410166}, which is based on the second data 
release (DR2) of the \sdss\ \protect\citep{2004AJ....128..502A}. It contains 
\sersic\ fits for 28089 galaxies in the redshift range $0.0033<z<0.05$.
For this paper, we use the VAGC elliptical aperture \sersic\ fits for
estimates of $r_{AB}$-band half-light radius, \sersic\ index and -flux; 
coupled with extrapolated circular aperture 
$u_{AB},g_{AB},r_{AB},i_{AB},z_{AB}$ fluxes.
The magnitudes were converted to absolute galactic foreground 
extinction-corrected magnitudes using the latest {\sc K-Correct} 
routines, which were also used for the original data 
\protect\citep{2003AJ....125.2348B}. 
We apply the following correction to convert the \sdss\ elliptical 
half-light $r_{AB}$-band sizes to rest-frame $V$-band (see 
$\S$~\ref{restsizes}): $R_e\left(V\right)=R_e\left(r\right)\times 1.011$. 
The redshift of the individual \sdss\ sources does not impact 
significantly on this correction factor. To obtain a rest-frame 
$B$-band size for the \sdss\ galaxies we use: 
$R_e\left(B\right)=R_e\left(V\right)\times 1.017$. 

We have chosen the VAGC rather than the fits to the magnitude--size and stellar
mass--size planes by \protect\citet{2003MNRAS.343..978S} for various reasons. Using 
the VAGC we have full control over all estimated parameters including
the photometric system, k-corrections, etc. Specificially, the fits by 
\protect\citet{2003MNRAS.343..978S} where performed on circularized size estimates
while we use elliptical \sersic\ measurements. The half-light
sizes and absolute magnitudes by \protect\citet{2003MNRAS.343..978S} were provided 
only in \sdss\ filters, necessitating the use of color transformations 
and of additional luminosity function convolutions in order to obtain 
mean values for the same selection and photometric system as the \gems\ 
data. Furthermore, the VAGC 
allows us to repeat the same analysis procedure that was also used for
the \gems\ data. Finally, the VAGC incorporates the latest version of the 
\sdss\ pipeline, leading to more robust \sersic\ indices, fainter apparent 
limiting magnitudes and fewer problems with deblending large sources. 
Since the VAGC and the data used by \protect\citet{2003MNRAS.343..978S} have 
$\sim20,000$ sources in common we could verify that the measured 
parameters broadly agree with each other.

The VAGC does not contain stellar masses. Therefore, 
we have used the prescription given in \protect\citet{2003ApJS..149..289B} to convert 
a $\left(g-r\right)_{AB}$ color into a \sdss\ $r_{AB}$-band 
stellar mass-to-light ratio: 
\begin{equation}
\log\left(\mathcal{M}/L_r\right)=-0.306+1.097\times\left(g-r\right)_{AB}-0.15. 
\end{equation}
We have applied a correction of $-$0.15 to convert to a Kroupa IMF, in accord
with our \gems\ stellar masses. The stellar mass was then obtained from the 
following relation: 
\begin{equation}
\log\left(\mathcal{M}\right)=\log\left(\mathcal{M}/L_r\right)
-0.4\times\left(M_{r,S}-r_\odot\right), 
\end{equation}
with the absolute rest-frame \sersic\ magnitude 
$M_{r,S}=r_S-5\log\left(D_L\right)-25$, the apparent rest-frame \sersic\ 
magnitude $r_S$, the luminosity distance $D_L$ and the 
absolute magnitude of the sun $r_\odot=4.67$ in \sdss\ $r_{AB}$. Calculating 
a stellar mass in the same fashion for the lowest redshift \gems\ 
galaxies and comparing this estimate with the SED-based masses 
\protect\citep{2004PhDT.........3B} reveals no apparent systematic offsets.

\subsection{Rest-Frame $V$-band Sizes\label{restsizes}}

\begin{figure}
\figurenum{2}
\begin{center}
\includegraphics[width=3.25in,keepaspectratio,clip]{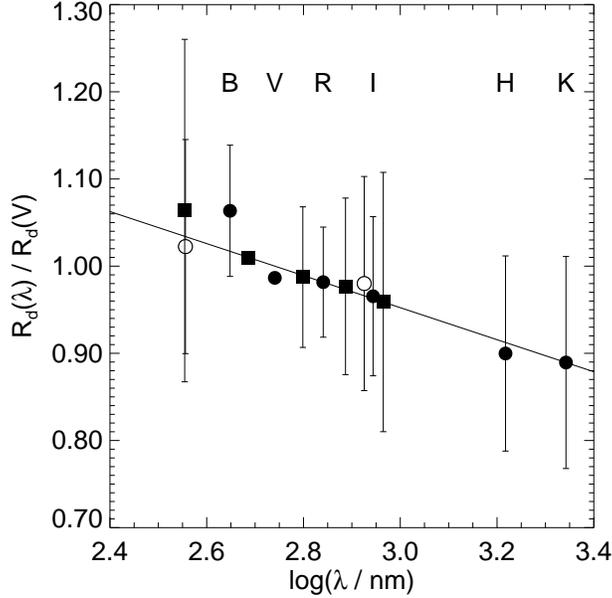}
\end{center}
\caption{The average ratio of the disk scale length
$R_d\left(\lambda\right)$ measured in various bands ($B$, $V$, $R$, 
$I$, $H$, $K$) over $R_d\left(V\right)$ measured in the $V$-band for 
the \protect\citet{1996A&AS..118..557D} data as a function of corresponding 
wavelength $\lambda$ (solid dots). 
The solid line marks a linear fit $f$, which 
is constrained to $f\left(2.74\right)=1$ at the $V$-band (not 
strictly requiring coincidence with the data point at the $V$-band). 
Over the redshift range sampled by the \gems\ data the size
corrections as inferred from this plot are of order $\pm3\%$. Errors
indicate the dispersion of the distribution of
$R_d\left(\lambda\right)/R_d\left(V\right)$; they do not represent 
errors of the mean values plotted here. Solid boxes 
mark data points from the \sdss\ DR2 data set (see $\S$~\ref{sdss} 
for details). Since there is no direct measurement in the $V$-band 
available for these objects, the data points are simultaneously 
fit minimizing the total offset between \sdss\ and de Jong 
values using the $g_{AB}$-band as the reference filter.
Open symbols represent 
measurements from the \gems\ survey where in a certain redshift bin 
one of the filters matched the rest-frame $V$-band. \label{fig_sizecorr}}
\end{figure}

Galaxies are known to exhibit radial color gradients. As a result of
this, galaxy sizes vary as a function of wavelength and the measured
physical size evolution of the galaxy population could be skewed by
the effects of band shifting with redshift. Therefore, we have not
simply converted our apparent half-light sizes $\reapp$ measured in
the F850LP filter to a physical value, but instead have applied a
color gradient correction to each individual galaxy according to its
redshift to correct the size to the rest-frame $V$-band. For a sample
of local galaxies, \protect\citet{1996A&AS..118..557D} presents the relative
disk scale lengths, which for a pure disk corresponds to
$R_e=1.678\times R_d$, in the $B$-, $V$-, $R$-, $I$-, $H$- and
$K$-bands. Figure~\ref{fig_sizecorr} illustrates this ratio of the
disk scale lengths in one band to the size measured in the $V$-band,
as a function of the corresponding wavelength. A linear fit with the
intercept fixed to 1 at the $V$-band results in a slope of
$a_R=-0.184$, corresponding to correction factors varying by only $\pm
3\%$ over the whole redshift range. All future references to effective
radii $R_e$ are to sizes corrected to the rest-frame $V$-band.

In order to obtain rest-frame sizes for the \sdss\ data we have
calculated the ratio of the circularized half-light sizes in the five
\sdss\ bands, divided by the size in the \sdss\ $g_{AB}$-band. We overplot
the resulting values in Fig.~\ref{fig_sizecorr}, minimizing in a
simultaneous fit the offset between the \sdss\ points and the other
$V$-band normalized measurements. The agreement between the various
measurements is striking. This supports the validity of the {\it
  average} correction to obtain rest-frame sizes, bearing in mind the
20\% galaxy-by-galaxy scatter, and that this method, strictly
speaking, applies only to nearby galaxies.

Given the possible rapid evolution of galaxy disks in the last 8
billion years, it is not inconcievable that the `average' disk color
gradient has evolved considerably since $z \sim 1$.  In a subsequent
paper we will reconstruct the rest-frame $B$-band for individual
galaxies and estimate sizes directly from this image to account for
this effect.  As an interim solution, we have tested the applicability
of the local average relation on distant galaxies in \gems .  We have
fit all \gems\ galaxies in the F606W band using exactly the same
approach used to fit in F850LP.  Owing to significant differences in
the depth of the F606W and F850LP data, and F606W's extra sensitivity
to ongoing star formation, we consider the F606W fits at this stage to
be preliminary\footnote{While many galaxy fits were reasonably
  successful, a non-negligible fraction of the fits are substantially
  in error.  Thus, while on average, the F606W fits are reliable, it
  is impossible at this stage to use a weighted sum of the F606W and
  F850LP fits to directly estimate the rest-frame $B$- or $V$-band
  sizes on a galaxy-by-galaxy basis.}.  From these fits we selected
those sources for which one of the bands corresponds to the rest-frame
$V$-band and measure the size ratio at $z\sim 0.08$ (F606W $\sim$
$V_{\textrm{\tiny{rest}}}$) and at $z\sim 0.64$ (F850LP $\sim$
$V_{\textrm{\tiny{rest}}}$). The average values from these
measurements are overplotted in Fig.~\ref{fig_sizecorr}. They confirm
the trend seen in the \protect\citet{1996A&AS..118..557D} and SDSS data,
supporting the validity of the correction we have applied to the
data\footnote{It is worth recalling that the corrections implied by
  this relation are rather small, $\sim 3$\% for the average
  \gems\ galaxy.  Furthermore, the evolution of average rest-frame
  $V$-band surface brightness is dominated by galaxies with $z \ga
  0.6$, where the F850LP samples rest-frame $V$-band almost directly,
  and by the SDSS data at $z \le 0.05$; thus, further reducing our
  sensitivity to any errors in the size correction.  }.

\subsection{Completeness\label{completeness}}

\begin{figure}
\figurenum{3}
\begin{center}
\includegraphics[width=3.25in,keepaspectratio,clip]{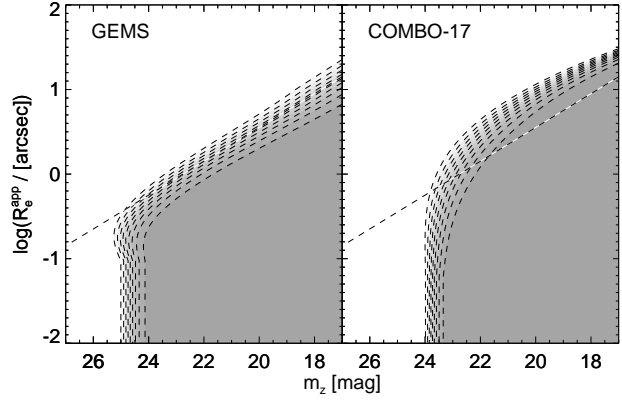}
\end{center}
\caption{The \gems\ (left panel) and the \combo\ (right panel) detection 
probabilites as a function of apparent $z$-band magnitude $m_z$ and apparent 
half-light radius $\reapp$. The contours indicate 
different detection probability levels (from light to dark shades of grey: 
10\%, 20\%, 30\%,...~90\%). For comparison, in both plots a line of 
constant apparent surface brightness $\mu_z^{\textrm{\tiny{app}}}=24$ mag
arcsec$^{-2}$ is shown.}
\label{fig_completeness}
\end{figure}

In order to estimate the limitations of the \gems\ survey we have
performed extensive simulations of artificial disk galaxy light profiles
(see H\"au\ss ler et al.~2005, in prep.). By inserting a number of such
artificial disk images with purely exponential profiles (\sersic\ index
$n=1$), and subsequently re-running the source detection and fitting 
process (including removal of bad fits according to $\S$~\ref{selection}), 
we calculate our success rate: the completeness as a function of apparent
effective radius $\reapp$ and apparent magnitude $m_z$. It turns out that
the contours of constant detection probability in the $\reapp$-$m_z$-plane
(see Fig.~\ref{fig_completeness}) lie along lines of constant apparent
surface brightness:
\begin{equation}
\mu_z^{\textrm{\tiny{app}}}=m_z+2.5\log(2\pi q)+
5\log(\reapp / \left[\textrm{arcsec}\right]),
\end{equation}
in the limit of bright magnitudes ($q$ is the axis ratio). At the faint
magnitude limit, however, the lines of constant detection probability are
at constant magnitude. The precise location of such a line depends also on
the axis ratio of the objects. In the absence of dust, an object with high
inclination has a higher detection probability than a source of the same
apparent magnitude but viewed face-on.

We model the detection probability as a function of the apparent
magnitude. A double exponential model provides a good fit to the data
(for a detailed description see appendix~\ref{appendix_completeness}).
Both the shape and the characteristic magnitude limit at which a specific
detection probability is reached depend on the apparent size and the axis
ratio.

Our final sample contains only the objects with redshift estimates from
\combo\ and therefore we must also account for the \combo\ completeness
limit. \protect\citet{2003A&A...401...73W} have calculated the completeness of
\combo\ as a function of apparent $R$-band aperture magnitude $\raper$,
redshift and $U-V$ color.  In order to show the \combo\ completeness 
contours on Figs.\ 3, 4, 6, 7 and 10, we statistically transform
the \combo\ completeness map into the $m_z - R_e$ plane (Appendix B).
We adopt this analytic approximation to the \combo\ completeness in 
the rest of this paper, but note that the 
use of either the true \combo\ completeness map or the analytical
mapping of the completeness maps onto the $m_z - R_e$ plane in the analysis
that follows does not affect our conclusions.

We combine the \gems\ detection probability and the \combo\ completeness
by multiplying the two values for each individual object:
\begin{equation}
p=p_{\textrm{\tiny{GEMS}}}\times p_{\textrm{\tiny{COMBO-17}}}.
\end{equation}
We can now estimate the combined detection probability $p$ of individual
galaxies. Since later on we weight galaxies by the inverse of the detection
probability we have taken special care when using very low detection
probability values. In order to avoid attributing large weights
to any given galaxy (which would then dominate the whole sample), we
remove any object with $p<5\%$ from the sample (a total of 14 sources). 
For the main analysis presented here,
we only include objects with a detection probability $p>50\%$. In appendix
\ref{appendix_analysis} we discuss in more detail how the detection
probability will impact on the evaluation of the data especially in the
magnitude--size plane, which is also the reason for not removing galaxies
with $0.05<p<0.5$ from the sample altogether. In
Fig.~\ref{fig_completeness_data} we illustrate the resulting detection
probability function in the $\reapp$-$m_z$-plane. 

\begin{figure}
\figurenum{5}
\begin{center}
\includegraphics[width=3.25in,keepaspectratio,clip]{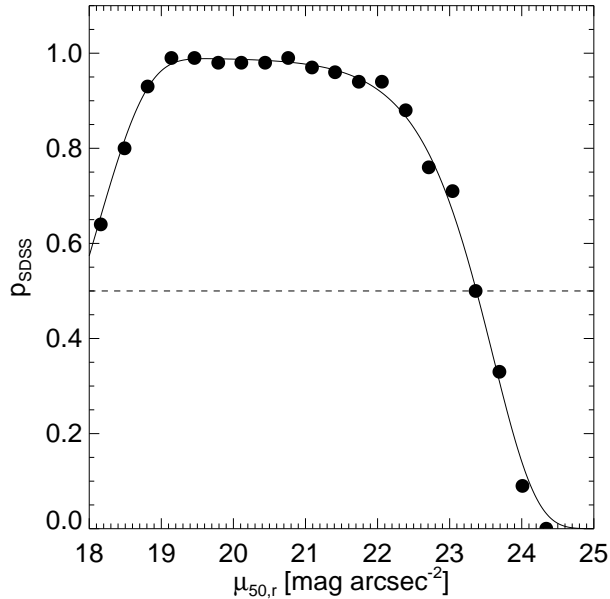}
\end{center}
\caption{The \sdss\ completness as a function of apparent surface 
brightness $\mu_{50,r}$. The dots indicate the actual surface brightness 
values as given in \protect\citet{astro-ph/0410166}. The solid line represents 
our analytical fit to that relation.}\label{fig_sdss_completeness}
\end{figure}

The completeness $p_{\textrm{\tiny{SDSS}}}$ of the \sdss\ data is
parameterized as a function of surface brightness $\mu_{50,r}$ and
position on the sky RA ($\alpha$) and dec ($\delta$)
\protect\citep{astro-ph/0410166}:
$p_{\textrm{\tiny{SDSS}}}\left(\mu_{50,r},\alpha,\delta\right)=
f_{ti}\left(\mu_{50,r}\right)\times f_{sp}\left(\mu_{50,r}\right)
\times f_{ph}\left(\mu_{50,r}\right)\times
f_{got}\left(\alpha,\delta\right)$, where $f_{ti}$ is the ``tiling''
fraction, $f_{sp}$ is the spectroscopic completeness, $f_{ph}$ is the
photometric completeness and $f_{got}$ is the fraction of main targets
for which a classification was obtained in this object's sector, as
described in more detail in \protect\citet{astro-ph/0410166}. The resulting
completeness as a function of surface brightness we present in
Fig.~\ref{fig_sdss_completeness} for the case
$f_{got}\left(\alpha,\delta\right)=1$. Note that the rapid drop of the
completeness at high surface brightnesses directly results from the
improper deblending of the largest nearby galaxies.  We have
approximated the data points given in \protect\citet{astro-ph/0410166} with
the following analytical formula:
\begin{eqnarray}
p_{\textrm{\tiny{SDSS}}}&=&0.99\times\exp\left(-\exp\left(
\frac{\mu_{50,r}-23.6}{0.6}\right)\right)\nonumber\\
&&\times\left(1-\exp\left(-\exp\left(\frac{\mu_{50,r}-18.1}
{0.7}\right)\right)\right)
\end{eqnarray}
In the subsequent analysis we only consider objects with a completeness 
$p_{\textrm{\tiny{SDSS}}}\geq 0.5$, in order to match the selection 
of the \gems\ galaxies. 

\section{Analysis of Completeness and Selection Effects}\label{sec_analysis}

In the following sections we evaluate the magnitude--size and stellar
mass--size relations as a function of redshift. We have subdivided our
sample of disk galaxies into five redshift bins, each of which spans a
range of 0.2 in redshift, centred on $z=0.2,0.4,0.6,0.8,1.0$, plus an
additional redshift bin at $z\sim0.0$ for the \sdss\ data.

In Fig.~\ref{fig_completeness_data} we show the combined completeness map
with observed disks with $0.35<q<0.65$ overplotted. 
The galaxies in the sample form a
relatively tight relation in the apparent magnitude--size plane. Inspecting
the slope $\alpha$ of this relation one realizes that it is close to, but
not exactly equal to a line of constant surface brightness. A linear fit
provides a slope $\alpha\sim-0.15$. In physical quantities this slope
closely matches that of a line of constant volume density, i.e.~a law such
that the ratio of flux and the cube of the radius is constant
($\alpha=-0.1\overline{3}$), rather than a line of constant surface
density ($\alpha=-0.2$). The fact that the slope does not match that of a
constant surface brightness implies that measuring average surface
brightnesses depends to some extent on the range in magnitudes over which
the average is calculated. Thus, in order to quantify the
evolution of the surface brightness one has to make sure that the same
range of absolute magnitudes is observed at all redshifts.

The sample becomes approximately magnitude-limited at $m_z\sim23.5$. This
limit is imposed by the \combo\ redshifts; fainter galaxies cannot be
assigned reliable redshifts. Furthermore, we find no galaxies at brighter
magnitudes $m_z<23$ with detection probabilities less than 50\%. 
Since we show in Fig.~\ref{fig_completeness_data} galaxies of all
redshifts, this implies that {\it the \gems\ data are not limited in 
surface brightness at any redshift}, even at the highest bin.
Therefore, our subsequent analysis is not affected by a possible 
completeness-induced truncation of the surface brightness distribution
of the galaxy population at any redshift. We
conclude that the combined \gems\ + \combo\ sample is
essentially magnitude-limited only, with surface brightness playing a
minor role. This conclusion is robust to the detailed choice of axis
ratios. 

We have translated these completeness contours to the absolute magnitude--size
plane in Fig.~\ref{fig_abscomp}.  To estimate the absolute magnitude, 
we fit a third-order polynomial to the 
``average apparent $z$ minus rest-frame apparent $V$ color'' 
$\left\langle m_{z}-m_{V}^{\textrm{\tiny{rest}}}\right\rangle$ of our sample
as a function of redshift.  Obviously a redshift dependence cannot
fully model this color, leading to a small additional scatter 
of the data relative to the transformed completeness map\footnote{These
transformed completeness maps are not used in the science analysis;
rather, they are included in the figures for presentational 
purposes alone.}.

In Fig.~\ref{fig_abscomp} we also overplot the \sdss\ completeness. 
At the low surface 
brightness edge a fairly large number of \sdss\ objects are found 
with very low completeness values;  
the VAGC does not sample the full distribution of 
surface brightnesses.  We adopt an absolute magnitude
cut of $M_V<-20$ in this paper: brighter than this limit
the size distribution is sufficiently narrow that the full range
of surface brightnesses is well-sampled.  In order 
to estimate where the apparent magnitude limit $m_r^{\lim}=17.77$ starts 
to affect the galaxy distribution we convert $m_r^{\lim}$ into an 
absolute magnitude $M_V^{\lim}=-18.8$ for the highest redshift in the 
VAGC using a color transformation for a typical Sbc
\protect\citep{1995PASP..107..945F}.  Again, this limit is fainter
than our adopted absolute magnitude cut.

Inspection of Fig.\ \ref{fig_abscomp} shows that the sample
reaches $M_V < -20$ in the highest redshift bin; therefore
in what follows we restrict our analysis to this absolute magnitude
range at all redshifts.   This selection
leaves 3584, 76, 176, 704, 671 and 559 disk galaxies in the respective 
redshift bins $z=0.0,0.2,0.4,0.6,0.8,1.0$; a total of $3584+2186$ galaxies. 
This magnitude cut implies that our results are
applicable only over this brightness range.  

We have explored in detail the influence on the average surface
brightnesses and surface densities of varying the 
$p > 50$\% criterion, the surface brightness range over which 
one averages, and the absolute magnitude range considered.  
The influence of the $p$ cut is negligible; the surface brightness
and magnitude ranges do affect the average surface brightnesses/densities,
and great care must be taken to choose appropriate integration ranges.
These issues are discussed where relevant in \S 5, and 
in great detail in Appendix \ref{appendix_analysis}.

\section{Analysis of the magnitude--size and 
Stellar mass--size Relation}\label{sec_mag_mass_size}

For our subsequent analysis we define the absolute rest-frame effective
surface brightness in the $V$-band as:
\begin{equation}
\mu_V=M_V+5\log R_e+2.5\log q+38.568,\label{eq_murest}
\end{equation}
with the absolute rest-frame magnitude in the $V$-band from \combo\ and
the half-light radius $R_e$ in kpc. The constant 38.568 results from using
sizes in kpc and luminosity distances in Mpc. Note that this formula is
correct even for a general \sersic\ profile. In the analysis of the 
evolution of $\mu_V$ we will only address the bright galaxy population 
with $M_V<M_V^{\lim}=-20$. Moreover, we define the ``equivalent'' absolute 
rest-frame surface mass density
\begin{equation}
\log\Sigma_\mathcal{M}=\log\mathcal{M}-2\log R_e-\log\left(2\pi q\right),
\label{eq_rhorest}
\end{equation}
where the SED-estimated stellar galaxy mass $\mathcal{M}$ is given in
$\mathcal{M}_\odot$. In the case of $\log\Sigma_\mathcal{M}$ we
restrict the sample to galaxies with $\logm >
\log\mathcal{M}^{\lim}=10$. We calculate average values of the surface
brightness $\muav$ and the surface mass density $\rhoav$, correcting
for incompleteness by weighting indiviual galaxies by the inverse of
their detection probability as a function of redshift. We obtain
errors on the estimated mean values by performing an extensive
Monte-Carlo analysis (see appendix~\ref{appendix_analysis}).

\subsection{The Magnitude--Size Relation}\label{sec_mag_size}

In Fig.~\ref{fig_magsize} we present the magnitude--size relation for
disk galaxies in six redshift bins extending to $z\sim1.1$. We stress
that the completeness contours shown in the figure are only indicative
as they were calculated for a fixed axis ratio $q=0.5$ and the central
redshift of the corresponding bin (see also
Fig.~\ref{fig_completeness}). Therefore, especially in the $z=0.2$
redshift bin, we see many galaxies ``spilling over'' into the
incompleteness regions, which is a result of the non-negligible range
of $M_V$ cutoffs over redshifts $0.1<z<0.3$. To illustrate this effect
we overplot vertical lines corresponding to an apparent magnitude
$m_z=24$ at the centre, low and high end of each redshift bin (for
$z\geq 0.2$).  With increasing redshift (co-moving volume) the spread
of the completeness becomes smaller. The detection probabilities for
individual galaxies, however, were calculated according to their exact
magnitude, size and axis ratio and not relative to the plotted
completeness contours. In the case of the $z\sim0.0$ redshift bin we
only indicate the brightness level, below which the highest redshift
galaxies are not fully sampled.

\begin{figure}
\figurenum{8}
\begin{center}
\includegraphics[width=3.25in,keepaspectratio,clip]{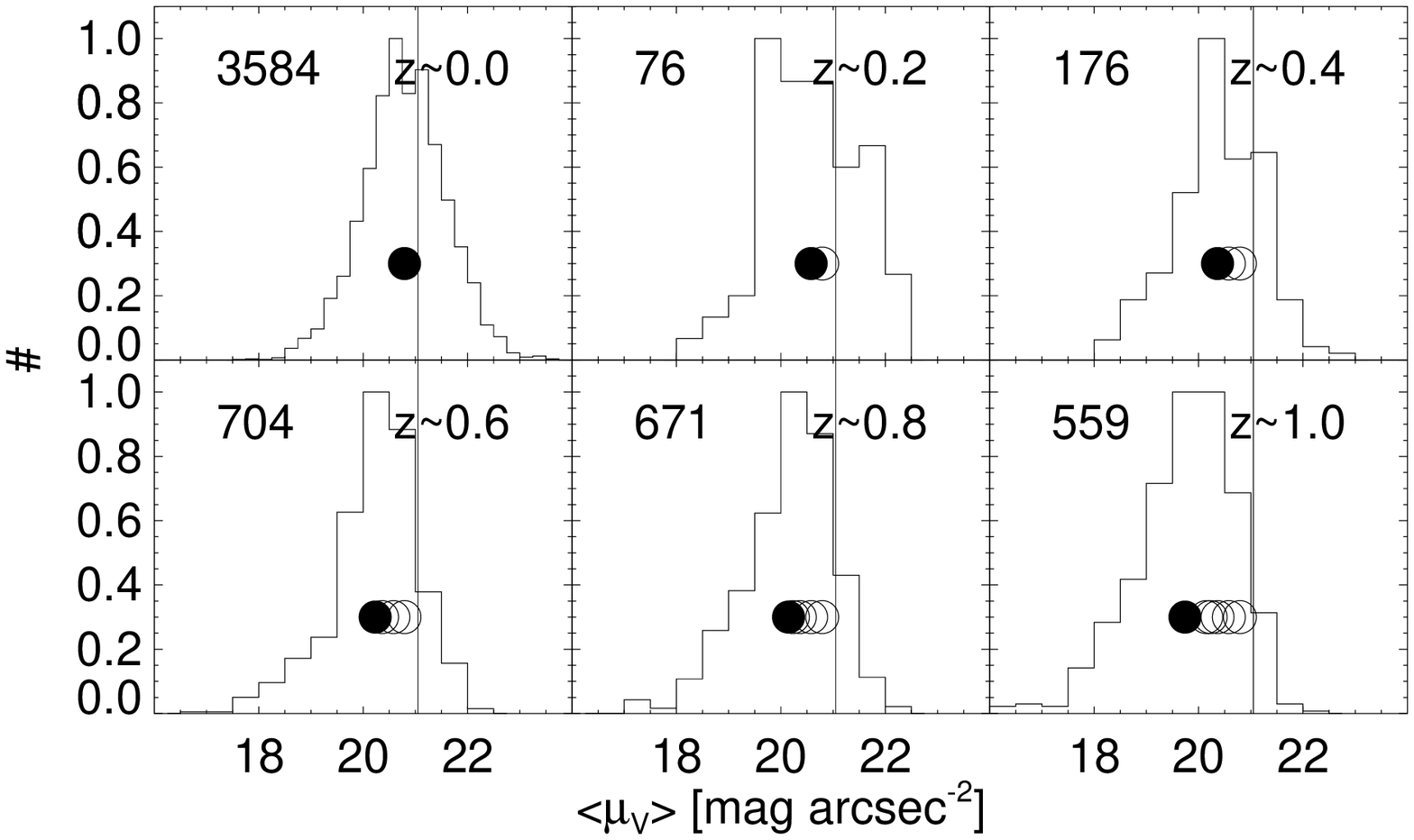}
\end{center}
\caption{Histograms of the absolute rest-frame surface brightness as a
function of redshift ($z\sim0$ from \sdss; $0.2<z<1.0$ from \gems). 
The histograms include weighting of individual
objects according to their completeness (y-axis peak normalized). Only
objects with a completeness exceeding 50\% and with $M_V<-20$ were
included in order to minimize the impact of selection effects. The black
circle marks the estimated mean value $\muav$ for each redshift bin. In each
redshift bin with $z>0.0$ we overplotted as open circles values of $\muav$
of all lower redshift bins to visualize the evolution in surface
brightness. The vertical line in each panel indicates the
\protect\citet{1970ApJ...160..811F} surface brightness converted to the $V$-band
\protect\citep[conversion given in][]{1995PASP..107..945F}. The numbers in the 
top left of each panel shows the number of sources used in each panel 
(not including weighting).\label{fig_sbhist}}
\end{figure}

As the completeness function limits us to detecting only the bright
galaxies at high redshift, we limit our analysis to galaxies with
$M_V < -20$.  Recall also that 
we have demonstrated in $\S$\ref{sec_analysis} 
and Appendix~\ref{appendix_analysis} that we are not 
limited in absolute surface brightness even
at the highest redshifts. Therefore, to evaluate the evolution of disk galaxies in
the magnitude--size plane we have calculated the average rest-frame
absolute surface brightness $\muav$ as a function of redshift including
weighting of individual galaxies according to their detection
probabilities. In Fig.~\ref{fig_sbhist} we show the weighted histograms of
$\mu_V\left(z\right)$ for each redshift bin. Indicated in each panel (at
each redshift bin $z_0$) are the estimated mean surface brightnesses
$\left\langle\mu_V\left(z_0\right)\right\rangle$ together with the mean
values of the preceeding redshift bins
$\left\langle\mu_V\left(z<z_0\right)\right\rangle$ for comparison. This
plot demonstrates clearly that there is a significant trend of increasing
surface brightness with increasing redshift.

\begin{figure}
\figurenum{9}
\begin{center}
\includegraphics[width=3.25in,keepaspectratio,clip]{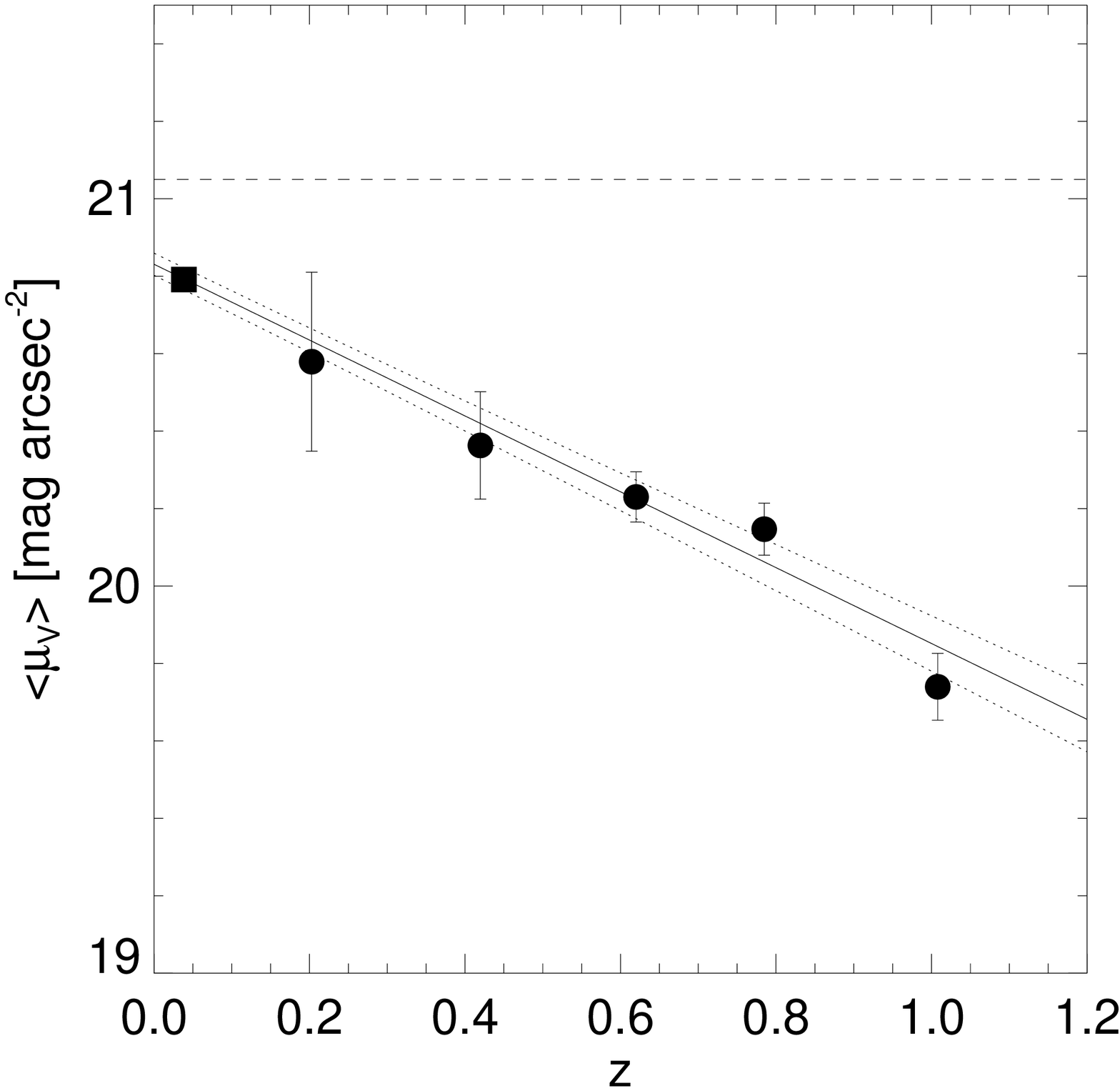}
\end{center}
\caption{Evolution of the average absolute surface brightness $\muav$.
Solid dots show the \gems\ data for the individual redshift bins; the
box symbol at $z\sim0.05$ indicates the \sdss\ data point. The error 
bars mark $2\sigma$ statistical errors. The horizontal 
line at $\mu_V=21.05$ represents the \protect\citet{1970ApJ...160..811F} surface 
brightness converted to the $V$-band. The solid and dotted lines mark a 
linear fit to all data points plus the $1\sigma$ confidence limits,
respectively.\label{fig_sbevo}}
\end{figure}

We demonstrate how the mean surface brightness of 
the disk galaxy population evolves by plotting $\muav$ as
a function of redshift in Fig.~\ref{fig_sbevo}. Fitting a linear function
to the data we find an intercept and slope of
$\left\langle\mu_V\left(z=0\right)\right\rangle=20.84\pm 0.03$ mag
arcsec$^{-2}$ and $\textrm{d}\left\langle\mu_V\left(z\right)\right\rangle/
\textrm{d}z=-0.99\pm 0.06$, respectively, 
thus an evolution of $\sim$1~mag to $z=1$.

\subsection{The Stellar Mass--Size Relation}

These stellar mass
estimates allow us to investigate the evolution of the analogous quantity
to the magnitude--size relation: the stellar mass--size relation. Working in
terms of stellar mass is useful not only because it is one step closer to
the quantities actually predicted by theory, but also because it removes
the evolution that is simply due to the aging of the stellar populations.
We present the stellar mass--size relation in Fig.~\ref{fig_masssize}.
Again, iso-density contours show the distribution of galaxies in the
$R_e$-$\mathcal{M}$ rest-frame plane. We use the same method as before to
correct the size estimates to the rest frame $V$-band (it is important to
note that ideally we would prefer to study stellar mass vs \emph{stellar
mass weighted size}, but we do not attempt this further correction here).
As in the case of the average surface brightness
$\left\langle\mu\left(z\right)\right\rangle$ we estimated the average
stellar surface mass density $\rhoav$, as defined in eq.~\ref{eq_rhorest},
for each redshift bin.

We found that, as in the case with the surface brightness, the
distribution of galaxies in the stellar mass--size plane does not fall
exactly along a line of constant stellar surface mass density, but is of
somewhat shallower slope. However, here the effect is much less pronounced
(also due to the width of the distribution) and therefore, the precise
cut-off in stellar mass, which is the equivalent of absolute magnitude, is
not as important.

Plotting mass as a function of magnitude for different redshift bins we
find that $\logm=10$ is a good approximation of the limiting mass in the 
highest redshift bin. In the calculation of $\rhoav$ we include the effects 
of completeness in exactly the same way as before, i.e.~we compute $\rhoav$ 
using a cut in stellar mass $\logm\geq10$ and we weight galaxies with the 
detection probabilities derived from Fig.~\ref{fig_completeness}.

\begin{figure}
\figurenum{11}
\begin{center}
\includegraphics[width=3.25in,keepaspectratio,clip]{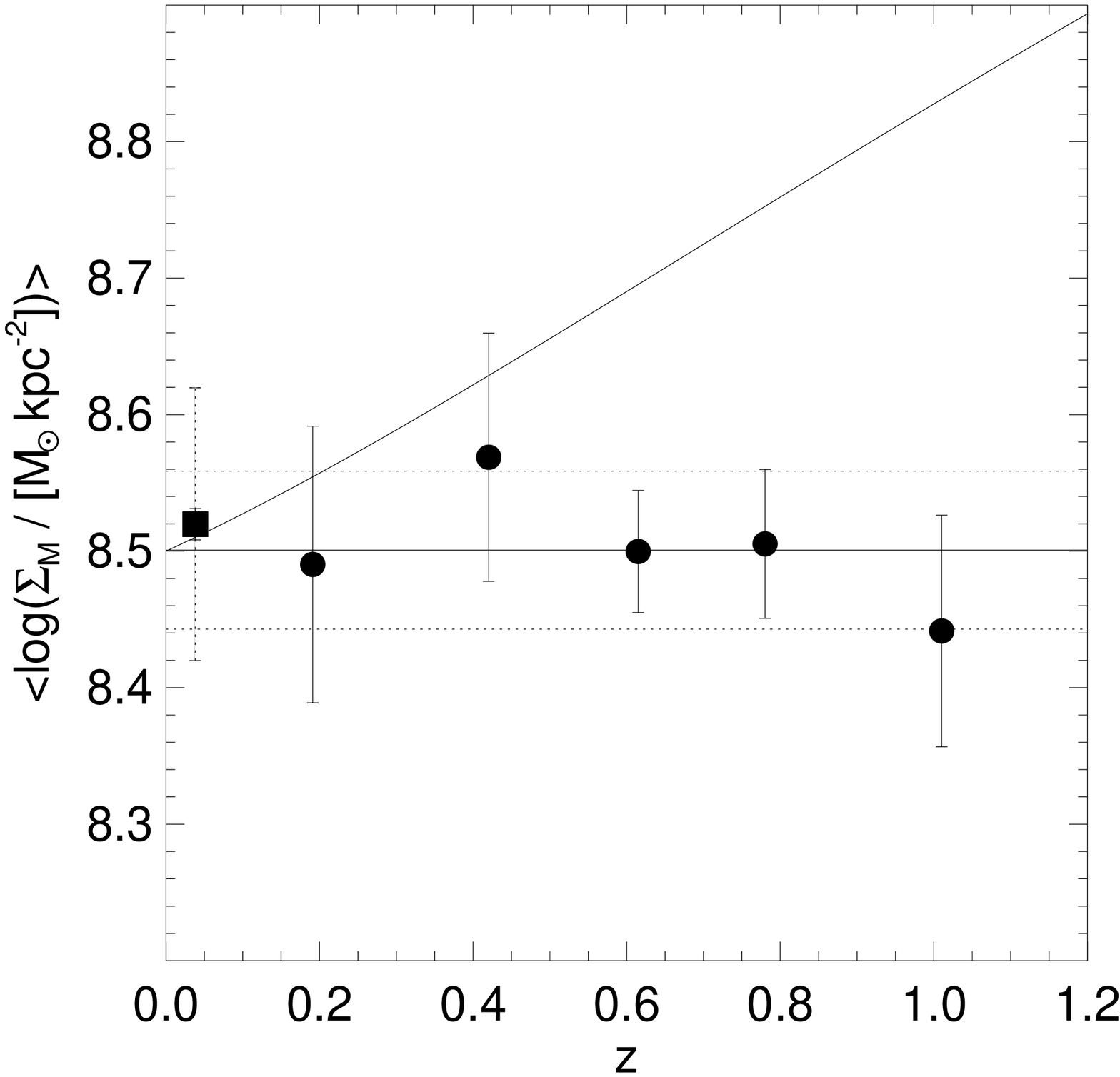}
\end{center}
\caption{Evolution of the average stellar surface mass density $\rhoav$.
Solid dots show the \gems\ data for the individual redshift bins; the
box symbol at $z\sim0.05$ indicates the \sdss\ data point. The solid error
bars indicate the $2\sigma$ statistical errors; the dotted error bar for 
the \sdss\ data point marks the $2\sigma$ systematic error resulting 
from the conversion of a mass-to-light ratio. The horizontal solid and 
dotted lines at $\rhoav=8.50\pm0.03$ represent a linear fit to the data 
with a constant slope of zero and the $1\sigma$ confidence limits, 
respectively. The diagonal line indicates the evolution as obtained from 
\protect\citet{1998MNRAS.295..319M}.\label{fig_massevo}}
\end{figure}

\begin{figure}
\figurenum{12}
\begin{center}
\includegraphics[width=3.25in,keepaspectratio,clip]{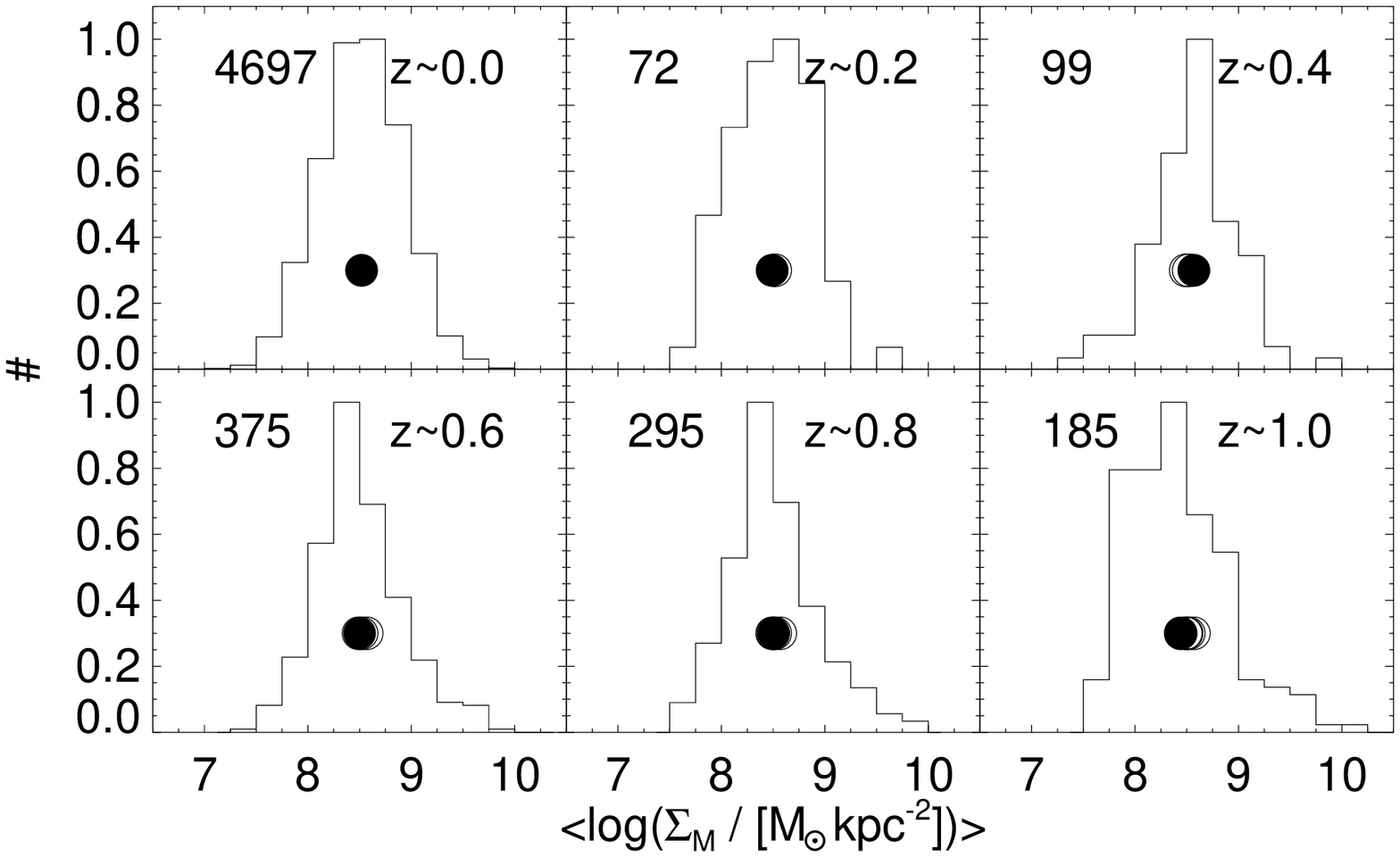}
\end{center}
\caption{Histograms of the stellar surface mass density as a function of
redshift ($z\sim0$ from \sdss; $0.2<z<1.0$ from \gems). 
The histograms include weighting of individual objects according
to their completeness (y-axis peak normalized). Only objects with a
completeness exceeding 50\% and with $\logm>\log\mathcal{M}^{\lim}=10$ 
were included in order to
minimize the impact of selection effects. The black circle marks the
estimated mean value $\rhoav$ for each redshift bin. In each redshift bin
with $z>0.0$ we overplotted as open circles values of $\rhoav$ of all
lower redshift bins to visualize the evolution in surface mass
density. The numbers in the 
top left of each panel shows the number of sources used in each panel 
(not including weighting).\label{fig_rhohist}}
\end{figure}

In Fig.~\ref{fig_massevo} we plot $\rhoav$ as a function of redshift and
find that the average surface mass density, to first order, does not
evolve significantly with redshift. The overall data values are found
within
$8.44<\left\langle\log\Sigma_{\mathcal{M}}\left(z\right)\right\rangle<8.57$.
This is also illustrated in Fig.~\ref{fig_rhohist} where we plot the 
histograms of the stellar surface mass density for the individual redshift 
bins. The deviation of the lowest and the highest 
data point corresponds to only 34\% in surface mass density.
Fitting a line with constant slope zero to the data yields
$\left\langle\log\Sigma_{\mathcal{M}}\left(z\right)\right\rangle =8.50\pm
0.03$.  We stress that the validity of this estimate does depend
strongly on systematic errors in the measurement of the stellar masses.
The error bars do not account for such effects and therefore might present
a somewhat oversimplied view. 

The constancy of the stellar mass--size relation above $\logm\sim 10$
since $z \sim 1$ comprises a strong constraint on models of disk
galaxy evolution.  The simplest possible interpretation of the data is
that galaxies grow inside-out: assuming that galaxies can only
increase their stellar mass with time, in order to stay on the
stellar-mass size relation as they grow in mass, galaxies must
increase their scale-lengths accordingly.  Yet, clearly, more complex
and physically-motivated models will also be capable of fitting the
data.

\section{discussion}\label{discussion}

\subsection{Surface Brightness Evolution}

In order to facilitate comparison with previous studies, we repeat the 
analysis in the rest-frame $B$-band (using absolute
$B$-band magnitudes from \combo\ and correcting the sizes to $B$-band). 
We convert the effective surface brightnesses
to central surface brightnesses assuming effective size and disk scale
length scale as $R_e=1.678\times R_d$:
\begin{equation}
\mu_{0,B}=M_B+5\log R_e-5\log(1.678)+38.568,\label{eq_mu0}
\end{equation}
This is strictly true only for pure disk galaxies, but should be a
reasonable approximation since the peak of our \sersic\ index
distribution roughly coincides with the exponential case $n\sim 1$. As
before, we find strong evolution in the rest $B$-band surface brightness
with redshift. For the intercept and slope in the rest-frame $B$-band
we find $\left\langle\mu_{0,B}\left(z=0\right)\right\rangle=21.11\pm
0.03$ mag arcsec$^{-2}$ and $-1.43\pm 0.07$, respectively (see
Fig.~\ref{fig_sbevo_B}).

In contrast to this picture of strong evolution, several previous
authors have found results consistent with weak or no evolution in the
average surface brightness out to $z\sim1$
\protect\citep[e.g.,][]{1999ApJ...519..563S,2004ApJ...604L...9R}. In this
section, we discuss how these apparently contradictory findings, based
on similar data, can be reconciled.

\subsubsection{Are the Datasets Significantly Different?}

We can rule out differences in the datasets as the source of our
divergent conclusions.  Owing to the similarity of the datasets, we
can reproduce the analysis of \protect\citet{2004ApJ...604L...9R} in some
detail.  \protect\citet{2004ApJ...604L...9R} assessed the average $B$-band
central surface brightness of their sample as a function of redshift,
limited in surface brightness to $\mu_{0,B}^{\lim} < 20.6$.  For both
the \goods\ and the \gems\ data sets the \protect\citet{2004ApJ...604L...9R}
surface brightness limit implies removing half to two thirds of all
galaxies at $0.25<z<0.50$ that are detected above the absolute
magnitude limit and have a measured redshift. Note that only $\sim5\%$
of all galaxies were excluded at the highest redshift
($1.00<z<1.25$). Obviously, by using only one third of galaxies with
the highest surface brightness, one introduces a strong bias in the
measurement of $\left\langle\mu_{0,B}\left(z\right)\right\rangle$ and
the derived value will therefore not represent the average properties
of disk galaxies at that redshift. Adoption of the surface brightness
limit used by \protect\citet{2004ApJ...604L...9R} yields very consistent
results to theirs for
$\left\langle\mu_{0,B}\left(z\right)\right\rangle$ (right-hand panel
of Fig.\ 13). For the \gems\ data we find evolution at less than the
0.4 mag arcsec$^{-2}$ level using their selection criteria. As
expected the high redshift data points are the least affected by their
surface brightness limit. However, at lower redshift the results
achieved using their selection criteria start to deviate
systematically from the analysis we presented earlier. Specifically,
the lowest redshift point with the surface brightness cut is more than
10$\sigma$ off the expected value (as estimated from our linear
relation) without such a cut.  \protect\citet{1999ApJ...519..563S} adopted a
very similar strategy, and also found very weak evolution, although in
their case low number statistics are also an important source of
uncertainty (there are only 5 and 6 galaxies in their lowest two
redshift bins, respectively).

\subsubsection{Are the Analysis Techniques Different?}

We argue that the divergence between our conclusions and those of
\protect\citet{1999ApJ...519..563S} and \protect\citet{2004ApJ...604L...9R} is 
driven primarily by important differences in the analysis techniques.

The analysis of \protect\citet{1999ApJ...519..563S} and
\protect\citet{2004ApJ...604L...9R}, justifiably, imposed the selection
function of high-redshift galaxies on the low-redshift galaxy
population, and asked whether the average surface brightness of
galaxies which one could have in principle seen at $z \sim 1$ has
evolved.  Clearly, because of cosmological surface brightness dimming
the bulk of nearby galaxies would be invisible if placed at $z \sim
1$, and are omitted from consideration. One then finds little
difference in the population of local galaxies that would be
observable at $z\sim1$.


In this paper, we adopt a different approach.  In essence, we step
gradually outwards from low redshift to higher redshift, asking at
each stage if there is any evidence that the results are significantly
biased due to cosmological surface brightness dimming.  The $z \sim 0$
SDSS data are clearly not surface brightness limited for galaxies with
$M_V < -20$.  Stepping outwards to $z \sim 0.2$ in the GEMS data,
the surface brightness limits are well clear of the observed
drop-off in galaxy number density for galaxies with $M_V < -20$.
Similarly for $z \sim 0.4$, $z \sim 0.6$, and $z \sim 0.8$; at each
redshift we have clearly detected both sides of the size distribution
in a region where completeness is $> 90$\%, and the observed drop-off
is real.  Thus, the observed evolution, at least out to $z \sim 0.8$,
is a genuine property of the entire disk galaxy population and is
unaffected by surface brightness dimming.  

At $z \sim 1$, it is less
obvious that the data are well clear of the selection boundaries --- we
correct for incompleteness using the estimates obtained by applying
our pipelines to artificial galaxies.  
Yet, even at the $z \sim 1$ bin we reach well beyond the peak of the 
surface brightness distribution (see Fig.~\ref{fig_sbhist}), within the 
limits that we can confidently correct for incompleteness. Therefore, 
either the evolution we measure in that bin is roughly correct, or 
the galaxy surface brightness distribution would have to be bimodal. In 
that case we could not observe a hypothetical second peak of low 
surface brightness galaxies. Furthermore, these galaxies would have to 
fade significantly (and faster than the ``normal'' galaxy population) 
with time, because otherwise we would detect these objects at lower 
redshifts. So far there are neither observational nor theoretical 
grounds on which to expect such a population of low surface brightness 
galaxies.

It is worth noting that if the galaxy
population did not evolve towards higher surface brightness at higher
redshift, we would have seen that in the data, as the sample out 
to $z \sim 0.8$ is clearly deep enough to probe the
entire $M_V < -20$ galaxy distribution in the high completeness
region.

\subsection{A New Population of High Surface Brightness Galaxies at High
Redshift?}

\protect\citet{1999ApJ...519..563S} and \protect\citet{2004ApJ...604L...9R} suggest that
at $z\sim 1$ a distinct population of very high surface brightness
galaxies emerges that is not detected at lower redshifts.
\protect\citet{1999ApJ...519..563S} describe these objects as sources with very
high surface brightness $\mu_{0,B}\lesssim18$ (they found 9 candidates; 18\% of
all galaxies detected at that redshift). \protect\citet{2004ApJ...604L...9R}
delineate this group of objects as compact ($R_e<0.8$ kpc) and bright
($M_V<-21.5$). Using their classification, $<5\%$ of the galaxies at
$z\sim 1$ fall into this category.

One might conjecture that the introduction of a new population of high
surface brightness galaxies at $z \sim 1$ simply arises in order to
interpret the increasing average surface brightness within a global
picture of a non-evolving
$\left\langle\mu_{0,B}\left(z\right)\right\rangle$.  Our results suggest that the
\emph{whole distribution} of surface brightnesses shifts with redshift,
naturally leading to a larger number of high surface brightness galaxies at
higher redshift. Furthermore, we find no evidence that the surface
brightness distribution changes its shape (at the 10\% level, 
from inspection of Fig.\ 8).

\begin{figure}
\figurenum{13}
\begin{center}
\includegraphics[width=3.25in,keepaspectratio,clip]{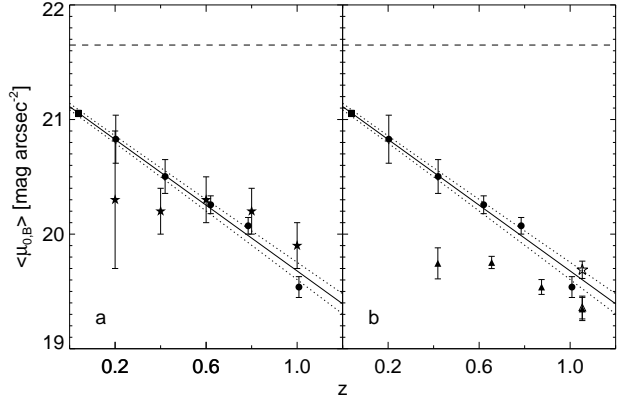}
\end{center}
\caption{Comparison of the evolution of the average absolute central
$B$-band surface brightness
$\left\langle\mu_{0,B}\left(z\right)\right\rangle$ with the literature. 
Solid dots show the \gems\ data for the individual redshift bins; the
box symbol at $z\sim0.05$ indicates the \sdss\ data point. The error 
bars mark $2\sigma$ statistical errors. The horizontal line at
$\mu_{0,B}=21.65$ represents the \protect\citet{1970ApJ...160..811F} surface
brightness. The solid and dotted lines mark a linear fit to
all data points (including the \sdss\ value) plus the $1\sigma$ confidence
limits, respectively. a) Comparison with 
\protect\citet[][star symbols]{1999ApJ...519..563S}. b) Reproducing the
\protect\citet{2004ApJ...604L...9R} analysis. Triangles show values obtained
from the \gems\ data using the selection in limiting surface brightness
and absolute magnitude as chosen by \protect\citet{2004ApJ...604L...9R}.
The open symbols indicate the impact of a population of high surface
brightness galaxies at the highest redshift bin using the definition of
\protect\citeauthor{2004ApJ...604L...9R} (open triangle, just barely visible above
the filled triangle) and the definition by
\protect\citeauthor{1999ApJ...519..563S} (open star). Especially when using the
latter definition, the \gems\ data with the \protect\citeauthor{2004ApJ...604L...9R}
selection are consistent with no surface brightness evolution.
\label{fig_sbevo_B}}
\end{figure}

Interestingly, both \protect\citet{1999ApJ...519..563S} and
\protect\citet{2004ApJ...604L...9R} introduce the appearance of this new group of
objects just at the high redshift limits of their surveys. At those
redshifts $z\sim 1-1.2$ their values for the average surface brightness
are generally in agreement with the \gems\ data points. We have shown in
Fig.~\ref{fig_sbevo_B}b that we can reproduce the effect of a flattening in
the evolution of $\left\langle\mu_{0,B}\left(z\right)\right\rangle$ by
introducing a hard upper surface brightness cut. However, the bulk of the
remaining evolution appeared in the highest redshift bin (and hence one
could propose the introduction of a new class of high surface brightness
galaxies to account for this). After removing the highest surface
brightness galaxies as classified by \protect\citet{2004ApJ...604L...9R} or
\protect\citet{1999ApJ...519..563S}, even the \gems\ data do not show a 
significant redshift-dependent trend 
in $\left\langle\mu_{0,B}\left(z\right)\right\rangle$ (see
Fig.~\ref{fig_sbevo_B}b). Although this line of reasoning appears to be
consistent it nevertheless has a major drawback. At the lowest redshift
the results should agree with the average surface brightness obtained from
the \sdss. This fact alone should raise
strong concerns regarding the global sampling of the local galaxy
population. Only strong evolution of
$\left\langle\mu_{0,B}\left(z\right)\right\rangle \sim 21.1 - 1.43z$ can account
for both convergence with the local data point and the high redshift 
results from \protect\citet{1999ApJ...519..563S}, \protect\citet{2004ApJ...604L...9R} and 
the results presented in this paper.

To summarize, we believe that the weak surface brightness evolution
found by \protect\citet{2004ApJ...604L...9R} and \protect\citet{1999ApJ...519..563S},
and the emergence of a `new population' of high surface brightness
galaxies at $z \sim 1$ results from differences between their analysis
technique --- which imposes the high-redshift selection function on
galaxies at all redshifts --- and our analysis technique, which
implicitly steps out gradually from the local towards the high
redshift universe, asking if there is any evidence for the galaxy
distribution running into the surface brightness detection limits.
Applying the same selection criteria as \protect\citet{2004ApJ...604L...9R},
we also found weak surface brightness evolution. The disadvantage of
this approach is that it cannot yield quantitative statements about
the evolution of the global ensemble of disk galaxies, especially at low
redshift.

\subsection{Comparison with Theoretical Expectations}\label{sec_theory}

The basic picture of disk formation within a hierarchical universe
posits that the dark matter and gas are `spun up' by tidal torques in
the early universe. The internal angular momentum is generally
characterized by the dimensionless spin parameter, $\lambda$.
Assuming that the gas does not suffer significant loss of specific
angular momentum during collapse, the size of the resulting disk $R_d$
is expected to scale as $R_d \propto \lambda r_i$, where $r_i$ is the
radius enclosing the gas before collapse 
\protect\citep[see e.g.][]{1998MNRAS.295..319M}. 
In cosmological N-body simulations, it
is found that the distribution of values of $\lambda$ for dark matter
halos follows a characteristic log-normal form, and that the value of
$\lambda$ does not correlate with halo mass, nor does the distribution
of $\lambda$ evolve with time \protect\citep{bullock:spin}. Thus, to first
order, we expect the size of a disk of fixed mass to scale with time
in proportion to the virial radius of the dark matter halo:
\begin{equation}
R_d\left(z\right)=R_d\left(0\right)\times\left[\frac{H\left(z\right)}
{H\left(0\right)}\right]^{-2/3},
\end{equation}
where $R_d\left(0\right)$ is the scale length at $z=0$ and $H(z)$ is
the Hubble parameter as a function of redshift
\protect\citep{1998MNRAS.295..319M}. Using the definition of the surface mass
density eq.~\ref{eq_rhorest} we find:
\begin{equation}
\log\Sigma_\mathcal{M}\left(z\right)=\log\Sigma_\mathcal{M}\left(0\right)
+\frac{4}{3}\log\frac{H\left(z\right)}{H\left(0\right)},
\end{equation}
with the surface mass density at redshift zero $\log\Sigma_\mathcal{M}
\left(0\right)$. Since we are interested in the relative evolution
only, we normalize the curve to the observed value
$\log\Sigma_\mathcal{M}\left(0\right)=8.5$ and show the redshift
dependence in Fig.~\ref{fig_massevo}. The expectation of this very
naive model is that disks at $z\sim1$ should be a factor of two denser
at fixed mass than they are at the present day, in clear contradiction
with the observational results.

In reality, however, we expect there to be several other competing
factors. For example, the internal density profile of the dark matter
halo, as commonly characterized by the concentration $c$, will also
impact the final size of the disk, in the sense that halos with higher
concentration will produce smaller, denser disks. The average
concentration at fixed halo mass is a function of epoch, scaling as $c
\propto (1+z)^{-1}$ \protect\citep{bullock:profiles}. Thus, the fact that
halos were \emph{less} concentrated at $z\sim 1$ by about a factor of
two will tend to counteract the strong evolution in surface density
indicated above. As well, there are numerous other complications:
there is certainly not a straightforward relationship between halo
mass and the mass of baryons that collapse to form a disk; the
specific angular momentum of the baryons that comprise the disk may
not be equal to that of the dark matter halo; the disk size can be
affected by the presence of a pre-existing bulge; and halos with low
spin parameters and/or large disk masses may not be able to support a
stable disk \protect\citep{1998MNRAS.295..319M}. In addition, a proper
comparison of the predicted evolution of the disk mass-size relation
with the data requires a careful treatment of the observational
selection effects. We defer this analysis to a future work (Somerville
et al. in prep).

\section{Summary}\label{sec_summary}

Based on two-dimensional fits to the light profiles of all \gems\ sources
we have compiled a complete and unbiased sample of disk galaxies. Our disk
sample was defined by its radial profile, specificially by 
\sersic\ profiles with concentrations
lower than $n=2.5$. \combo\ provided us with redshifts, rest-frame
absolute magnitudes and stellar masses for $\sim 5700$ sources. In order
to compare the \gems\ data to a local reference we have obtained the VAGC,
containing the same information as provided by \gems\ for $\sim28,000$ 
nearby ($z<0.05$) \sdss\ galaxies. Inspecting the magnitude--size and 
the stellar mass--size relation for disk galaxies as
a function of redshift we have come to the following conclusions:

$\bullet$ At high redshifts $z\sim 1$ the \gems\ survey is complete
only for galaxies with absolute magnitudes $M_V\lesssim-20$ or stellar
masses $\logm\gtrsim10$. In order to properly address the potentially
severe biases that arise when one attempts to explore the evolution of
the galaxy population over this redshift range, we have computed a
detailed 2-dimensional selection function and introduced a lower
limiting absolute magnitude cut.

$\bullet$ Treating completeness and selection effects carefully, we find
that the average surface brightness of disk galaxies increases with
redshift, by about 1 magnitude from $z\sim1$ to the present in the
rest-frame $V$-band. 

$\bullet$ The values calculated in our study are consistent at the high
redshift end with the results of \protect\citet{2004ApJ...604L...9R} and
\protect\citet{1999ApJ...519..563S} and at the low redshift end with the value 
estimated from the \sdss\ VAGC. We have shown
that the reasons the studies of \protect\citet{1999ApJ...519..563S} and
\protect\citet{2004ApJ...604L...9R} reached rather different conclusions from our
own (weak or no surface brightness evolution over the same redshift range)
are primarily related to the way the data were analyzed, as well as to
problems with small number statistics in the lower redshift bins. In
particular, applying a hard lower surface brightness cut leads to removing
substantial numbers of galaxies in the low redshift bins, and to a strong
bias in the estimated value of the {\it average} surface brightness. This
approach yields average surface brightness estimates at low redshift
$z\sim0.2$--0.4 that do not converge with the ``zero redshift'' results
from SDSS. We confirmed that when we apply the same selection criteria
to the \gems\ data, we obtain results that are consistent with those of
\protect\citet{2004ApJ...604L...9R}.

$\bullet$ In contrast to the conclusions of \protect\citet{1999ApJ...519..563S}
and \protect\citet{2004ApJ...604L...9R}, we find that there is no need to appeal
to a new population of high surface brightness galaxies, which makes its
appearance at high redshift. The increased number of high surface
brightness galaxies at high redshift is a natural result of the surface
brightness evolution that we have detected.

$\bullet$ While the magnitude--size relation shows strong evolution with
redshift, we show that the stellar mass--size relation stays 
constant with time.  

$\bullet$ The most naive theoretical expectation is that disks of
fixed mass should be about a factor of two denser at $z\sim1$, in
clear contradiction with our results. Several competing factors
probably conspire to produce the weaker evolution that we observe.

$\bullet$ As the stellar mass of galaxies increases with time, the fact
that the surface mass density does not evolve as a function of redshift
implies that \emph{on average} disk galaxies form inside-out, 
i.e.~through increasing their disk scale lengths with time as they grow 
in mass.

\acknowledgments

Based on observations taken with the NASA/ESA {\it Hubble Space
Telescope}, which is operated by the Association of Universities for
Research in Astronomy, Inc.\ (AURA) under NASA contract NAS5-26555.
Support for the \gems\ project was provided by NASA through grant number
GO-9500 from the Space Telescope Science Institute, which is operated by
the Association of Universities for Research in Astronomy, Inc., for NASA
under contract NAS5-26555. EFB and SFS acknowledge financial support
provided through the European Community's Human Potential Program under
contract HPRN-CT-2002-00316, SISCO (EFB) and HPRN-CT-2002-00305, Euro3D
RTN (SFS). CW was supported by a PPARC Advanced Fellowship. SJ
acknowledges support from the National Aeronautics and Space
Administration (NASA) under LTSA Grant NAG5-13063 issued through the
Office of Space Science. DHM acknowledges support from the National
Aeronautics and Space Administration (NASA) under LTSA Grant NAG5-13102
issued through the Office of Space Science. CH and HWR acknowledge
financial support from GIF. KJ was supported by the german DLR 
under project number 50~OR~0404. 

Funding for the Sloan Digital Sky 
Survey (SDSS) has been provided by the Alfred P. Sloan Foundation, the 
Participating Institutions, the National Aeronautics and Space 
Administration, the National Science Foundation, the U.S. Department of 
Energy, the Japanese Monbukagakusho, and the Max Planck Society. The 
SDSS Web site is http://www.sdss.org/. The SDSS is managed by the 
Astrophysical Research Consortium (ARC) for the Participating 
Institutions. The Participating Institutions are The University of 
Chicago, Fermilab, the Institute for Advanced Study, the Japan 
Participation Group, The Johns Hopkins University, Los Alamos National 
Laboratory, the Max-Planck-Institute for Astronomy (MPIA), the 
Max-Planck-Institute for Astrophysics (MPA), New Mexico State 
University, University of Pittsburgh, Princeton University, the 
United States Naval Observatory, and the University of Washington.

\appendix

\section{Parameterization of the Detection Probability}
\label{appendix_completeness}

The detection probability $p$ for the \gems\ data as a function of apparent 
magnitude $m$ is well fitted by a double exponential function:
\begin{equation}
p_{\textrm{\tiny{GEMS}}}=\exp\left(-\exp\left(\frac{m-m_0}{\sigma}\right)\right)
\end{equation}
with the slope $\sigma$ and the characteristic magnitude $m_0$. Both the slope 
$\sigma=\sigma\left(\reapp,q\right)$ and the characteristic 
magnitude $m_0=m_0\left(\reapp,q\right)$ are a function of the apperant 
half-light radius $\reapp$ [pix] and the axis ratio $q$. Since the 
smallest objects included in the simulations have half-light radii 
$\reapp\geq 0.3$ pix we hold the \gems\ completeness fixed at 
sizes $\reapp<0.5$ pix: $\log R=\textrm{max}
\left[\log\left(\reapp\right),0.5\right]$. The slope $\sigma$ is defined as:
\begin{equation}
\sigma=\sigma_0\left(q\right)+\sigma_1\left(q\right)\times R
\end{equation}
with 
\begin{eqnarray}
\sigma_0\left(q\right)=0.0860+0.118\times q\nonumber\\
\sigma_1\left(q\right)=0.308-0.0634\times q
\end{eqnarray}
The characteristic magnitude $m_0$ is defined as:
\begin{eqnarray}
m_0&=&\textrm{min}[22.5+\tilde{m}\left(q\right)\times\log R
+\left(7.37-1.83\times\tilde{m}\left(q\right)\right)\times\log R^2
+\left(-3.44+0.60\times\tilde{m}\left(q\right)\right)\times\log R^3\nonumber\\
&&+\cos\left(7\log R-1.75\right)
\times\exp\left(-4.2\log R\right)
+5\log R+2.5\log\left(q\right),\mu_{\textrm{\tiny{max}}}\left(q\right)]
-5\log R-2.5\log\left(q\right)\nonumber\\
\end{eqnarray}
with 
\begin{eqnarray}
\tilde{m}\left(q\right)&=&5.325+5.373\times q-2.128\times q^2\nonumber\\
\mu_{\textrm{\tiny{max}}}\left(q\right)&=&29.80+0.0933\times q\nonumber\\
\end{eqnarray}

Similarly, we fit the detection probability for the \combo\ data by 
a double exponential function:
\begin{equation}
p_{\textrm{\tiny{COMBO-17}}}=n\times\exp\left(-\exp
\left(\frac{m-m_{0,c}}{\sigma_c}\right)\right)\label{pcombo}
\end{equation}
Since the \combo\ detection probability does not depend on the axis ratio, both
$\sigma_c=\sigma_c\left(\reapp\right)$ and $m_{0,c}=m_{0,c}\left(\reapp\right)$ 
take much simpler forms as functions of $\reapp$ only. The slope $\sigma_c$ is 
defined as:
\begin{eqnarray}
\sigma_c&=&0.168+0.388\times\nonumber\\
&&\exp\left(-\frac{1}{2}\left(
\frac{\min\left[-\log\frac{\reapp}{2},0\right]-2.131}{0.895}\right)^2\right)
\end{eqnarray}
and the characteristic magnitude $m_{0,c}$ is defined as:
\begin{equation}
m_{0,c}=23.85-0.274\times\tilde{R}+0.507\times\tilde{R}^2-0.403\times\tilde{R}^3
\end{equation}
with $\tilde{R}=\max\left[\log\reapp,0.3\right]$. The normalisation $n$ differs 
slightly from unity due to the effect of redshift focussing:
\begin{equation}
n=1.014+0.00112\times\tilde{R}
\end{equation}

\section{Incorporating the \protect\citet{2003A&A...401...73W} Completeness Map}
\label{appendix_combo_completeness}

The completeness map given in \protect\citet{2003A&A...401...73W} contains values for 
the \combo\ detection probability $p_{\textrm{\tiny{COMBO-17}}}=
p_{\textrm{\tiny{COMBO-17}}}\left(\raper,z,(U-V)_{\textrm{\tiny{rest}}}\right)$ 
as a function of the apparent $R$-band aperture magnitude $\raper$, 
the redshift $z$ and the $(U$-$V)_{\textrm{\tiny{rest}}}$ rest-frame color. In order 
to convert this completeness map into our magnitude--size frame 
$p_{\textrm{\tiny{COMBO-17}}}=p_{\textrm{\tiny{COMBO-17}}}\left(m_z,\reapp\right)$,
we take the following approach.

We start with a simulated catalogue containing a uniform
distribution of apparent $z$-band magnitudes, apparent half-light sizes
(uniformly distributed in $\log\reapp$) and redshifts. Then, 
we convert the \gems\ $z$-band magnitude $m_z$ 
into a \combo\ total $R$-band magnitude
$\rtot=\rtot\left(m_z,z\right)$.
The following polynomial fit to the data is an adequate
description (where $\textrm{RND}$ denotes a normally
distributed random number with a mean of zero and a standard deviation of 1):
\begin{eqnarray}
\rtot&=&m_z-[\textrm{RND}\times\left(-0.308+
0.0253\times m_z\right)\nonumber\\
&&-0.202-0.340\times z+3.984\times z^2-13.881\times z^3\nonumber\\
&&+13.918\times z^4-4.264\times z^5\nonumber\\
&&+2.951-0.376\times m_z+0.0110\times m_z^2]\label{eq1}
\end{eqnarray}

Next, we relate $\rtot$ to $\raper$ (the \combo\ completeness map
is expressed in terms of $\raper$).  
By assuming that the aperture loss for the disk galaxies in \combo\
is a function of the half-light radius. For the $n<2.5$ disk sample we
find a linear correlation between the difference of total and aperture
\combo\ magnitude $\rtot-\raper$ and $\sqrt{\reapp}$: 
\begin{equation}
\raper=\rtot-0.508+0.254\sqrt{\reapp}-0.226\times\textrm{RND}\label{eq2}
\end{equation}
The scatter about
this relation is only 0.23 mag.

Finally, we estimate the \combo\ $\left(U-V\right)$ color given the 
\gems\ $m_z$, redshift and size.  We find that 
the following description, which is a function of $m_z$ and redshift,
is an adequate representation of the data:
\begin{eqnarray}
\left(U-V\right)_{\textrm{\tiny{rest}}}&=&\textrm{RND}\times 0.270\nonumber\\
&&+0.480-0.534\times z+0.125\times z^2\nonumber\\
&&+2.417-0.107\times m_z\label{eq3}
\end{eqnarray}

Using these transformations, we assign a \combo\ detection probability
from the completeness map given in \protect\citet{2003A&A...401...73W} to 
each mock \gems\ galaxy, where the detection probability
is a function of $m_z$, $\reapp$ and $z$. We have compared
the results of the modeling to the direct values from the
\protect\citet{2003A&A...401...73W} completeness map. {\it Statistically the 
agreement is good and our subsequent conclusions are unaffected by 
which particular method is chosen.} We have carried out the analysis 
using both methods, arriving at the same conclusions. In the paper we 
refer to our statistical approach in order to clearly demonstrate the 
fact that \combo\ is somewhat deeper in terms of surface brightness.

\section[]{Analysis of Completeness and Selection Effects}
\label{appendix_analysis}

In order to address the effects of our completeness correction and sample
selection we take the following approach: For each redshift bin we
calculate histograms of $\mu_V\left(z\right)$ and
$\log\Sigma_\mathcal{M}\left(z\right)$ using the inverse detection 
probability $p$ of
each object as a weight. From these ``weighted'' histograms we measure
average values for the rest-frame absolute surface brightness in the
$V$-band, $\muav$ and stellar surface mass density, $\rhoav$.
These average values and the corresponding errors we obtain by
constructing 1,000 Monte-Carlo realisations of the \gems\ data for each
redshift bin. Each realisation consists of a random subsample of the 
whole data set containing as many sources as the full set, but allowing 
for duplicate data points.
The adopted average values originate from the average mean
value of the 1,000 simulations, while the error bars in
$\muav$ and $\rhoav$ were calculated from the
scatter of the 1,000 mean value estimations. Using such a procedure, we
are able to correct for galaxies missing in the \gems\ survey down to the
level where we can reliably estimate the detection probability $p$ when
calculating average mean values. 

The calculation of $\muav$ and $\rhoav$ is affected by three limitations.
At some limiting magnitude $m_z^{\textrm{\tiny{lim}}}$ the detection 
probability $p$ drops to zero. The same occurs at some limiting surface 
brightness $\mu_z^{\textrm{\tiny{app,lim}}}$. Both effects limit the 
range of absolute magnitude and surface brightness that is covered by the
\gems\ data. The higher the redshift, the brighter is the corresponding 
limiting absolute magnitude $M_V^{\textrm{\tiny{lim}}}$ and the limiting
rest-frame surface brightness $\mu_V^{\textrm{\tiny{lim}}}$. As a result of
this, we have to restrict the study of the average galaxy population at each 
redshift bin to the galaxies brighter than $M_V^{\textrm{\tiny{lim}}}$ and 
$\mu_V^{\textrm{\tiny{lim}}}$ corresponding to the highest redshift bin.
Finally, the value of $p^{\textrm{\tiny{lim}}}$, at which one does not 
include objects in the calculation of $\muav$ and $\rhoav$, also 
potentially impacts on the analysis. In the case of $\rhoav$ the absolute 
magnitude limit $M_V^{\textrm{\tiny{lim}}}$ translates into a limiting mass 
$\log\mathcal{M}^{\textrm{\tiny{lim}}}$.

As will be shown in our subsequent analysis we find 
$M_V^{\textrm{\tiny{lim}}}=-20$ and 
$\log\mathcal{M}^{\textrm{\tiny{lim}}}=10$. 
We will also provide further proof 
for the fact that the \gems\ data are not limited in surface brightness 
even at the highest redshift bin. Furthermore, we will show that our 
results are fairly independent of the choice of 
$p^{\textrm{\tiny{lim}}}$. We adopt a rather conservative value of 
$p^{\textrm{\tiny{lim}}}=0.5$. In order to demonstrate these results, we
calculate $\muav$ and $\rhoav$ for various combinations of 
$p^{\textrm{\tiny{lim}}}$, $\mu_V^{\textrm{\tiny{lim}}}$,
$M_V^{\textrm{\tiny{lim}}}$ and $\log\mathcal{M}^{\textrm{\tiny{lim}}}$. 

\begin{figure}
\figurenum{4}
\begin{center}
\includegraphics[width=5in,keepaspectratio,clip]{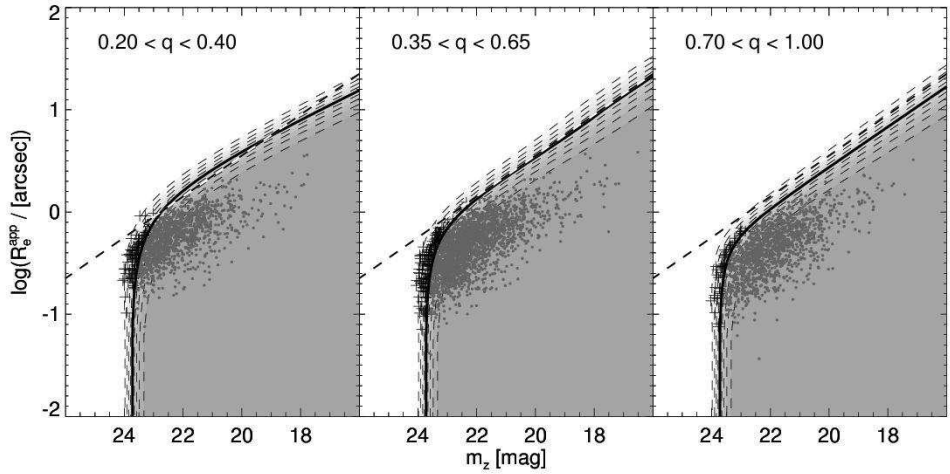}
\end{center}
\caption{The combined \gems\ / \combo\ completeness (dashed contours
indicating 10\%, 20\%, 30\%,...~90\%) in the
$\reapp$-$m_z$-plane. The solid contour shows the 50\%
completeness limit. Dots and pluses indicate disk galaxies with a
combined \gems\ detection / \combo\ redshift estimation 
probability $p>50\%$ and $p<50\%$, respectively. 
The panels show the completeness contours and data points for three 
different axis ratio ranges, indicated in the top of each plot.
The diagonal line indicates a constant apparent surface brightness 
$\mu_z^{\textrm{\tiny{app}}}=24$ mag arcsec$^{-2}$.
\label{fig_completeness_data}}
\end{figure}

\begin{figure*}
\figurenum{6}
\begin{center}
\includegraphics[width=5in,keepaspectratio,clip]{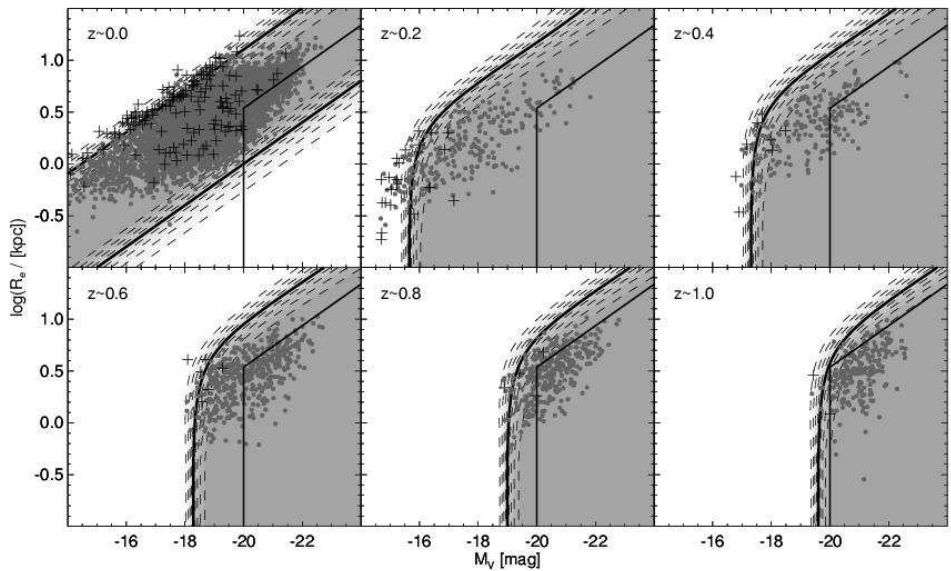}
\end{center}
\caption{The completeness as a function of redshift in the absolute
magnitude--size plane. The same completeness contours and objects from
the middle panel of Fig.~\ref{fig_completeness_data} are plotted. 
The $z\sim 0$ bin shows the \sdss\ data. The contours were computed
for the central redshift of each bin as indicated in the top left of each
panel. Galaxies with low detection probability are found along the
absolute magnitude limit, but not at bright magnitudes and faint surface
brightnesses. The box in each 
panel encloses a selection with $\mu_V^{\textrm{\tiny{lim}}}=22$ mag 
arcsec$^{-2}$ and $M_V^\textrm{\tiny{lim}}=-20$. This plot 
shows that the \gems\ data ($0.2<z<1.0$) are not limited in 
surface brightness. 
Galaxies are observed in regions where the completeness contours 
indicate detection probabilities $p<5\%$. This is a result of the finite
width of the redshift bins (see $\S$~\ref{sec_mag_size} and 
Fig.~\ref{fig_magsize} for further explanations).
\label{fig_abscomp}}
\end{figure*}

\begin{figure*}
\figurenum{7}
\begin{center}
\includegraphics[width=5in,keepaspectratio,clip]{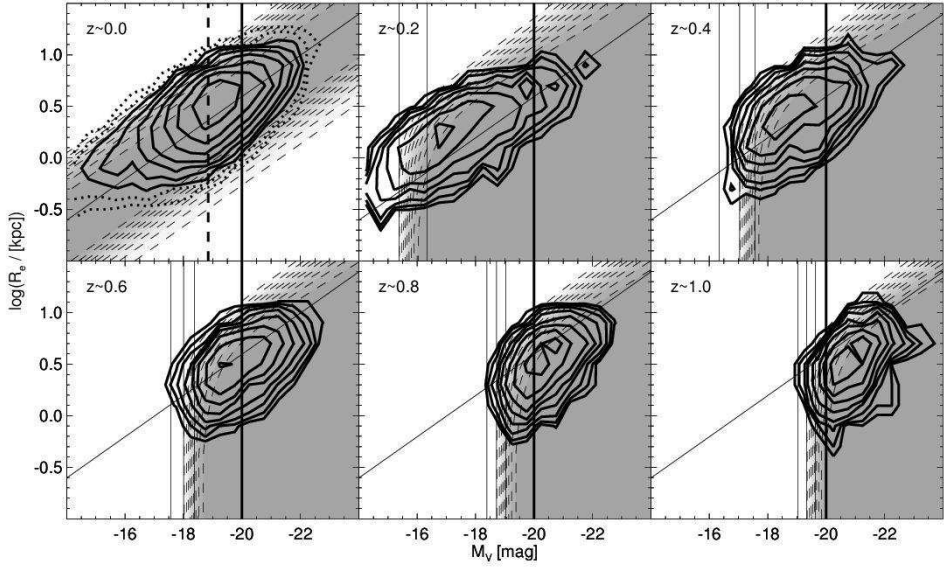}
\end{center}
\caption{The magnitude--size relation for six different redshift bins. The 
solid contours indicate 11 levels of fractional number of objects per 
unit area. To obtain the contours the density field was normalised by 
the total number of objects being plotted in each redshift bin and the 
following scaling was used: 
$\left(\exp\left(x/2\right)-1\right)\times0.0005$, with $x=[1,2,...,11]$. 
The dotted contours are only shown for the lowest redshift bin 
($z\sim0$; \sdss\ data) for being 
too noisy in the other bins. The grey-scale contours mark the same
completeness levels as shown in Fig.~\ref{fig_abscomp}. 
The thin vertical lines in each panel symbolise
an apparent magnitude $m_z=24$ converted to the rest-frame at three
different redshifts corresponding to the center, low and high end of each
redshift bin. This illustrates why apparently so many galaxies especially
in the low redshift bin were detected at completeness values close to
zero. The diagonal line in each panel corresponds to the average surface 
brightness $\mu_V=20.84$ mag arcsec$^{-2}$ in the lowest redshift bin. 
The thick vertical line represents the lower magnitude limit ($M_V=-20$), 
below which we exclude galaxies from the analysis. The thick dashed line 
shows the magnitude limit of the highest redshift \sdss\ data; at fainter
magnitudes the VAGC data become incomplete.\label{fig_magsize}}
\end{figure*}

\begin{figure*}
\figurenum{10}
\begin{center}
\includegraphics[width=5in,keepaspectratio,clip]{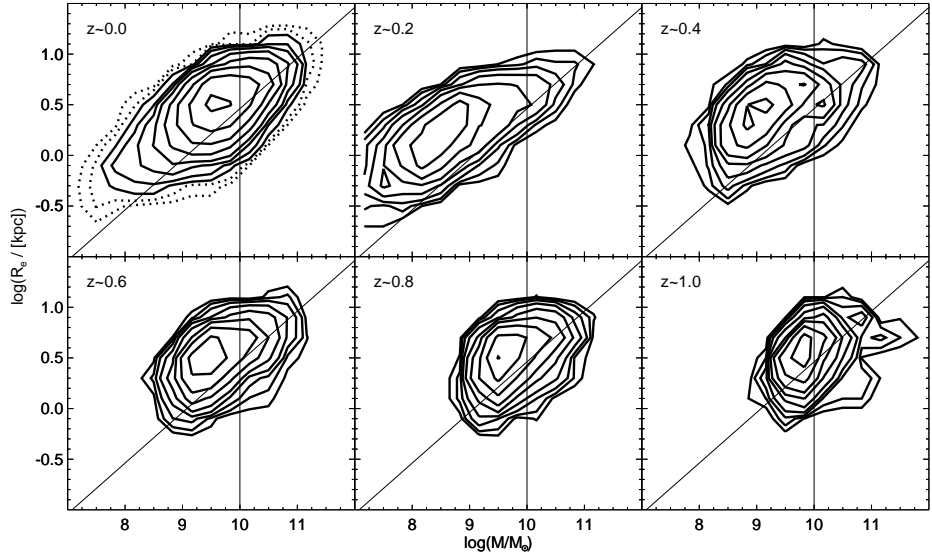}
\end{center}
\caption{The stellar mass--size relation for six different redshift bins 
($z\sim0$ from \sdss; $0.2<z<1.0$ from \gems).
The contours indicate levels of the same fractional number of objects per
unit area as in Fig.~\ref{fig_magsize}. The vertical line in each panel
marks $\logm=10$, which corresponds to the limiting stellar mass 
$\log\mathcal{M}^{\lim}$ applied to each redshift bin. The
diagonal line in each panel corresponds to the average surface mass
density measured from all redshift bins
$\left\langle\log\Sigma_{\mathcal{M}}\left(z=1\right)\right\rangle=8.50$.
\label{fig_masssize}}
\end{figure*}

\begin{figure}
\figurenum{C1}
\begin{center}
\includegraphics[width=3.25in,keepaspectratio,clip]{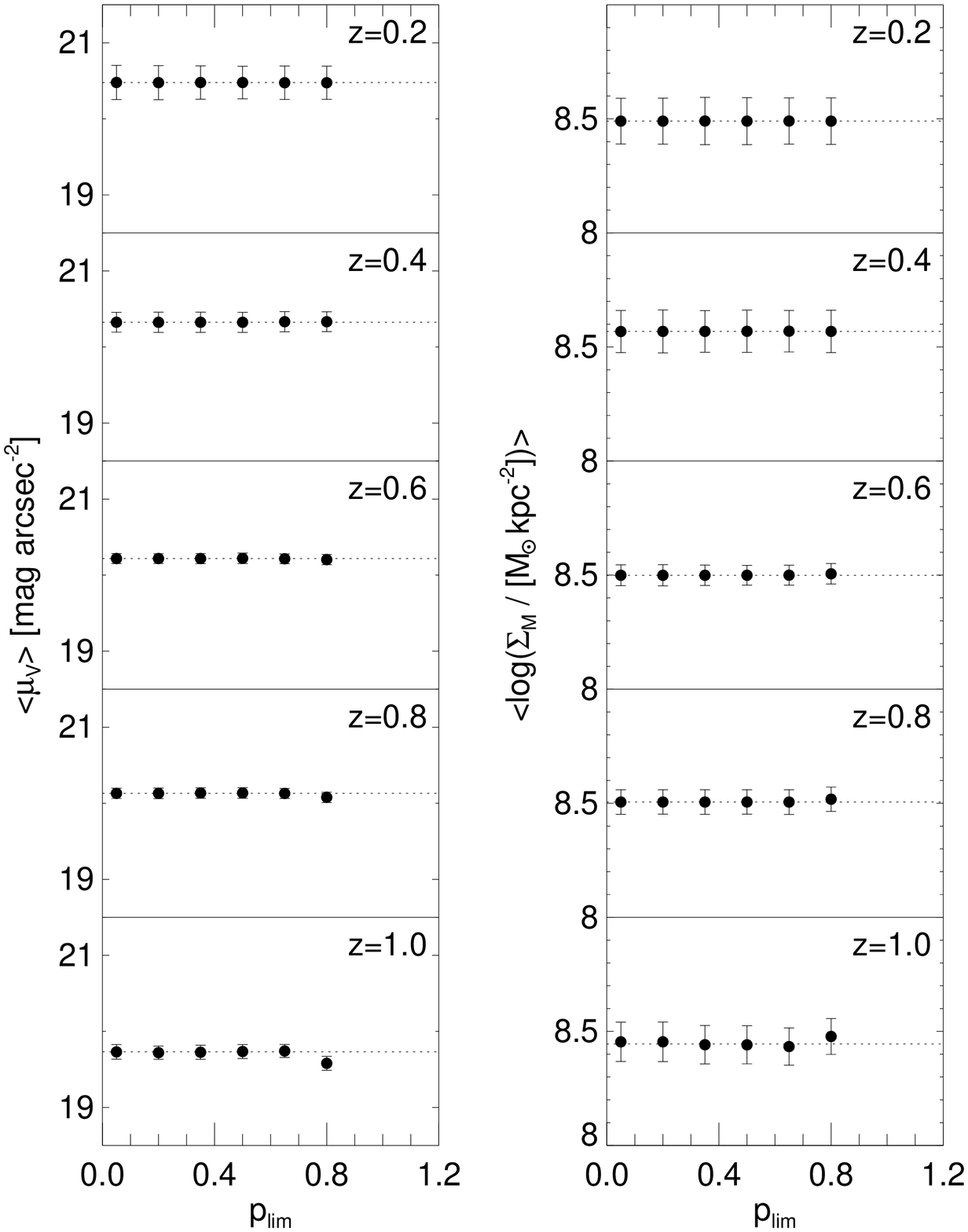}
\end{center}
\caption{The average surface brightness $\muav$ (left panel) and the
average surface mass density $\rhoav$ (right panel) for the
different redshift bins as a function of the adopted cut-off detection
probability $p_{\lim}$. Horizontal dotted lines mark the means of each
redshift bin. In both panels an absolute limiting magnitude
$M_V^{\lim}=-20$ was applied.\label{fig_comp}}
\end{figure}

In Fig.~\ref{fig_comp} we plot $\muav$ and $\rhoav$ as a function of 
the adopted $p^{\textrm{\tiny{lim}}}$ while holding 
$M_V^{\textrm{\tiny{lim}}}=-20$, 
$\log\mathcal{M}^{\textrm{\tiny{lim}}}=10$ and 
$\mu_V^{\textrm{\tiny{lim}}}=\infty$ 
constant. Both $\muav$ and $\rhoav$ do not vary significantly, i.e.~one
would obtain the same results for $\muav$ and $\rhoav$ using 
$p^{\textrm{\tiny{lim}}}=0.2$ or $p^{\textrm{\tiny{lim}}}=0.8$.
The reason for this is two-fold. On one hand the absolute magnitude limit 
$M_V^{\textrm{\tiny{lim}}}=-20$ and the lower stellar mass limit 
$\log\mathcal{M}^{\textrm{\tiny{lim}}}=10$ are chosen rather conservatively,
leading to a removal of almost all sources with $p<0.8$. On the other
hand, once the absolute magnitude limit reaches the region where the 
detection probability drops, galaxies fall along a line of constant 
apparent magnitude (see Fig.~\ref{fig_completeness_data}, at 
$m_z\sim 23.75$ the 50\% completeness contour is almost vertical).
Thus, calculation of a mean surface brightness (or surface mass density) 
is evenly (un-)affected by the completeness correction (independent of
surface brightness). Both these arguments arise from the fact that 
the \gems\ data is not limited in surface brightness.

\begin{figure}
\figurenum{C2}
\begin{center}
\includegraphics[width=3.25in,keepaspectratio,clip]{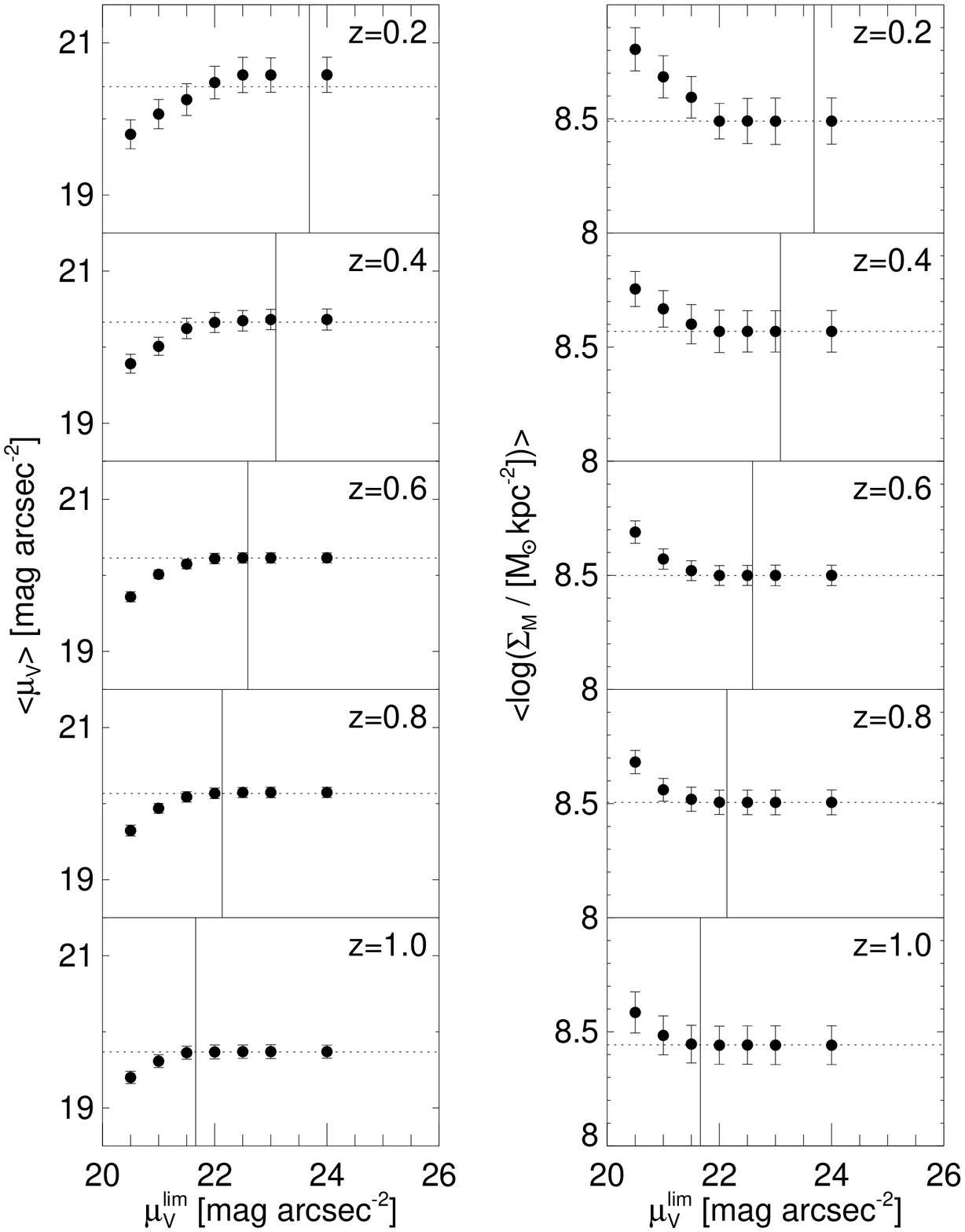}
\end{center}
\caption{The average surface brightness $\muav$ (left panel) and the
average surface mass density $\rhoav$ (right panel) for the
different redshift bins as a function of the adopted surface brightness
limit $\mu_{V}^{\lim}$. Horizontal dotted lines mark the means of each
redshift bin. The values at $\mu_{V}^{\lim}=24$ are plotted to demonstrate
the results in the case were no fixed surface brightness limit
($\mu_{V}^{\lim}=\infty$) is chosen. In both panels an absolute limiting
magnitude $M_V^{\lim}=-20$ and a detection probability cut $p_{\lim}=0.5$
was applied. The vertical lines correspond to the 50\% completeness 
limiting surface brightness for an axis ratio $q=0.5$ at the 
indicated redshift.\label{fig_mulim}}
\end{figure}

We repeat this exercise for $\mu_V^{\textrm{\tiny{lim}}}$
(see Fig.~\ref{fig_mulim}). This time we 
hold $p^{\textrm{\tiny{lim}}}=0.5$, $M_V^{\textrm{\tiny{lim}}}=-20$ 
and $\log\mathcal{M}^{\textrm{\tiny{lim}}}=10$ 
fixed. We find that there is a characteristic surface brightness at 
each redshift at which the estimated values of $\muav$ and $\rhoav$ 
systematically
start to deviate towards higher surface brightnesses. This has to be 
interpreted as the surface brightness at which one starts removing 
galaxies from the sample with the lowest surface brightness, thus 
shifting the average to higher surface brightnesses. Measuring
constant mean values at the lowest surface brightnesses, however, does
not necessarily imply that the average does not shift. It rather means
that we run into our completeness limit eventually, i.e.~we do not
detect the galaxy population at all that might yet exist at such a 
faint level. The question is, whether we reach a plateau in $\muav$ 
or $\rhoav$ before we run into the \gems\ surface brightness limit. 
To test this, we convert the apparent $z$-band surface 
brightness limit, i.e.~where a completeness level of $50\%$ is reached, 
as obtained from the \gems\ completeness map 
$\mu_z^{\textrm{\tiny{app,50\%}}}\sim23.9$ into a rest-frame 
surface brightness limit $\mu_V^{\textrm{\tiny{50\%}}}$ in the $V$-band
for each redshift bin using the following relation:
\begin{equation}
\mu_V^{\textrm{\tiny{50\%}}}=\mu_z^{\textrm{\tiny{app,50\%}}}+
\left[m_V^{\textrm{\tiny{rest}}}-m_z\right]-10\log\left(1+z\right)
\end{equation}
with a redshift $z$ dependent color term 
$m_V^{\textrm{\tiny{rest}}}-m_z$. The term $-10\log\left(1+z\right)$
arises from the surface brightness dimming $\propto\left(1+z\right)^4$,
which has to be accounted for when converting an apparent surface 
brightness to an absolute one. From a fit to the \gems\ data we 
obtain:
\begin{eqnarray}
m_V^{\textrm{\tiny{rest}}}-m_z&=&M_V-m_z+5\log D_L+25\nonumber\\
&=&0.562-0.111\times z+1.160\times z^2-0.841\times z^3
\end{eqnarray}
with the luminosity distance $D_L$. The values obtained in this manner
are indicated in Fig.~\ref{fig_mulim} as vertical lines. We find that 
we can calculate the average galaxy population representing values for
$\muav$ and $\rhoav$ at $z=0.2$ down to a limiting surface brightness
$\mu_V^{\textrm{\tiny{50\%}}}\sim23.7$. At higher redshift the 
corresponding value has dropped significantly, 
$\mu_V^{\textrm{\tiny{50\%}}}\sim21.7$ at $z=1.0$. Fortunately, even 
at that redshift we see that the average value $\muav$ has already 
flattened out, thus implying that even at the high redshift end of 
the \gems\ survey we do sample the full distribution of surface 
brightnesses. The same line of reasoning also applies to $\rhoav$.

\begin{figure}
\figurenum{C3}
\begin{center}
\includegraphics[width=3.25in,keepaspectratio,clip]{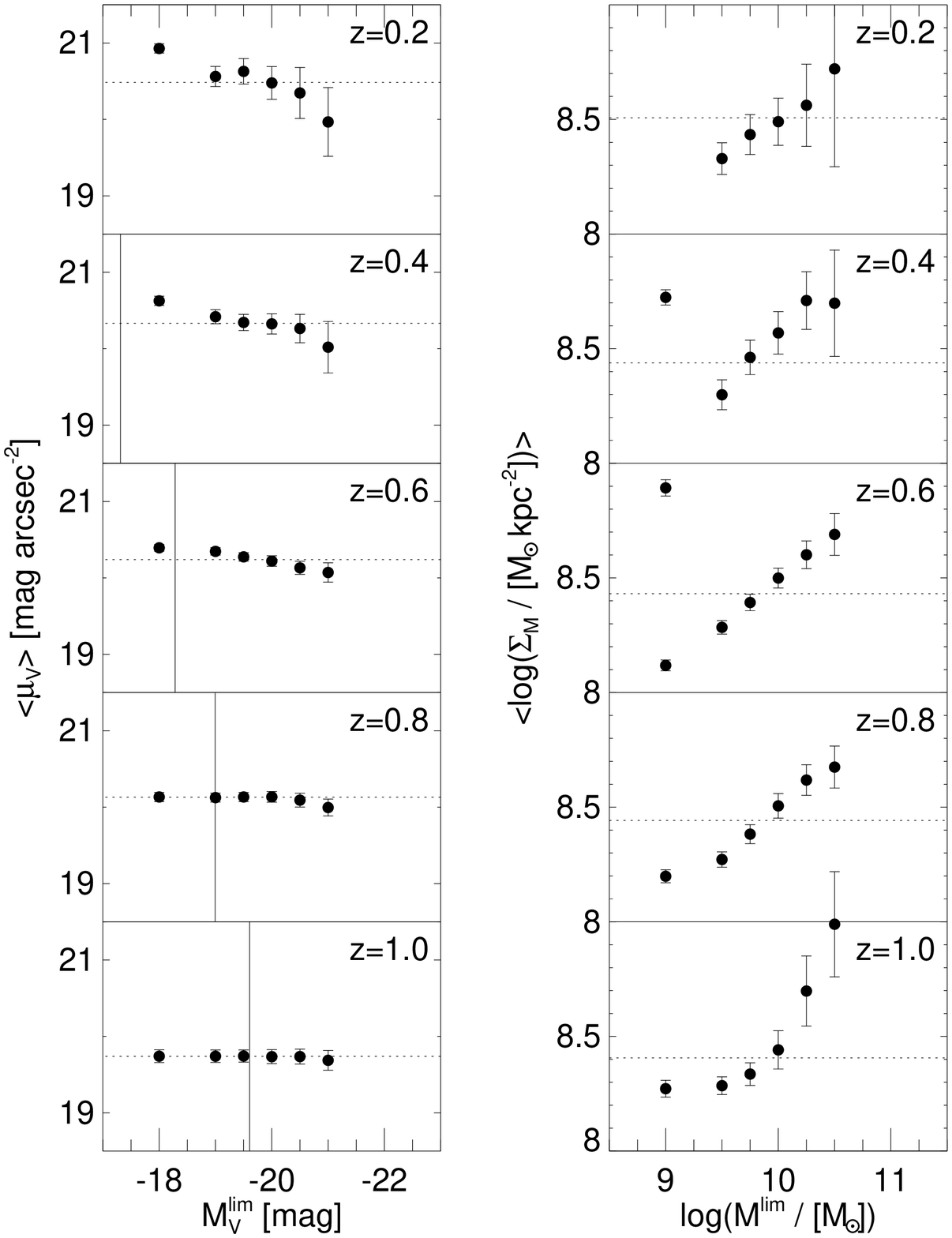}
\end{center}
\caption{The average surface brightness $\muav$ (left panel) and 
the average surface mass density $\rhoav$ (right panel) for the
different redshift bins as a function of the adopted absolute magnitude
limit $M_V^{\lim}$ or the limiting stellar mass $\log\mathcal{M}^{\lim}$ (left 
and right panel, respectively). Horizontal dotted lines mark the means of each
redshift bin. The values at $M_V^{\lim}=-18$ (left) and 
$\log\mathcal{M}^{\lim}=9$ (right) are plotted to demonstrate
the results if no fixed absolute magnitude limit ($M_V^{\lim}=-\infty$)
or stellar mass limit ($\log\mathcal{M}^{\lim}=-\infty$) is chosen. In both panels a detection probability cut $p_{\lim}=0.5$ was
applied. The vertical lines in the left panel correspond to the 50\% 
completeness limiting absolute magnitude at the indicated redshift.
\label{fig_mlim}}
\end{figure}

Finally, we examine the effect of the choice of 
$M_V^{\textrm{\tiny{lim}}}$ and $\log\mathcal{M}^{\textrm{\tiny{lim}}}$ 
while holding $p^{\textrm{\tiny{lim}}}=0.5$
and $\mu_V^{\textrm{\tiny{lim}}}=\infty$ fixed (see 
Fig.~\ref{fig_mlim}). Similarly to Fig.~\ref{fig_mulim} we overplot
the 50\% detection limit $m_z^{\textrm{\tiny{50\%}}}\sim23.7$ at the 
faint magnitude end of the completeness map converted to a rest-frame
absolute magnitude limit $M_V^{\textrm{\tiny{50\%}}}$. Now the
conversion reads:
\begin{eqnarray}
M_V^{\textrm{\tiny{50\%}}}&=&m_z^{\textrm{\tiny{app,50\%}}}+
\left[m_V^{\textrm{\tiny{rest}}}-m_z\right]-5\log D_L-25\nonumber\\
&=&m_z^{\textrm{\tiny{app,50\%}}}+
\left[M_V-m_z\right]
\end{eqnarray}
with the same definitions as above. We have not attempted to construct 
a similar relation for the case of $\log\mathcal{M}^{\textrm{\tiny{lim}}}$. 
We find that both $\muav$ and $\rhoav$ vary systematically as a function of 
$M_V^{\textrm{\tiny{lim}}}$ and $\log\mathcal{M}^{\textrm{\tiny{lim}}}$.
The reason for this is that the 
distribution of galaxies in the magnitude--size plane does not exactly
fall along a line of constant surface brightness, but has a slightly
steeper slope. This is most obvious at the lower redshift bins where
we have the largest dynamic range in absolute manitudes. Therefore,
our results are strictly true only for the adopted limiting magnitude
$M_V^{\textrm{\tiny{lim}}}=-20$. If one were to repeat our evaluation
with deeper data, thus reaching fainter absolute limiting 
magnitudes at the highest redshift, one should expect to find slightly
different absolute values for $\muav$ and $\rhoav$. However, if the 
distribution of galaxies does not change with redshift in the 
magnitude--size plane, which would imply differential evolution, one 
would measure the same relative differences. The same line of reasoning
of course also holds for $\log\mathcal{M}^{\textrm{\tiny{lim}}}=10$. 

The only way to 
circumvent this problem would be to move from measuring evolution
in the surface brightness to a new variable $\rho$, which matches the 
observed slope of the low redshift population\footnote{In fact, it 
should match the 
slope at all redshifts. However, we have found that usually at 
higher redshift the dynamic range in absolute magnitude is too small 
to reasonably constrain the slope.}. Fitting the slope 
in our lowest redshift bin, one reads off approximately
$\log\rho\propto M+2.5\times 3\log R$, with magnitude $M$ and radius 
$R$. This quantity has the physical 
dimensions of a volume density instead of a surface density, being
proportional to the radius cubed.

\clearpage

\clearpage

\end{document}